%% file: 4933SuppessedSUSYFINALX.tex
\font\caps=cmcsc10 at 12pt
\font\eightrm=cmr8 at 8pt
\documentclass{oxarticle}
\usepackage{amsmath, amssymb}
\usepackage{feynmf}

\input{Johnmacros21}
\newcounter{orange} 
\newcommand{\articlenumber}{4933SuppessedSUSYFINALX.tex}

\proofmodefalse
\usepackage{color} 

\begin{document}

\begin{center}

\vspace*{1in}

{ \huge Chiral SUSY Theories with a Suppressed SUSY Charge\\
[.5cm] }  
%

\vspace*{.1in}
%


\renewcommand{\thefootnote}{\fnsymbol{footnote}}

{\caps John A. Dixon\footnote{jadixg@gmail.com
} }\\[1cm] 

\end{center}

\vspace*{.2in}
 \begin{center}
 {\bf Abstract}
\end{center}

 The well-known Chiral and Gauge SUSY Actions realize the SUSY charge in terms of  transformations among the Fields. These transformations are included in the Master Equation by coupling them to Sources.   Here we show that there are new local SUSY Actions where the Chiral SUSY transformations are realized in terms of  transformations among both Fields and Sources. These Actions can be easily obtained from the Chiral case by  a very simple and local  `Exchange Transformation', which carries along all the interactions without difficulty. For these new SUSY Actions, the SUSY charge does not exist in the relevant sector, because Sources do not satisfy Equations of Motion.

Nevertheless, the `Exchange Transformation' ensures that the new Master Equation is  true for the new Action.  As a consequence,  the Master Equation also is true for the new 1PI Generating Functional.
This implies that a `Suppressed SUSY Charge' version of  SUSY is still present.  SUSY certainly becomes  more obscure and less constrained in this case. But it is still very restrictive. 

The new theories can be obtained from the old theories by using a special technique, but it is not true that they are a sort of `broken version of supersymmetry'.  They are simply a new type of theory that is governed by Supersymmetry, but without the use of Supercharges (except perhaps in some sectors). 

   In particular the number of physical Bosonic and  Fermionic degrees of Freedom are not equal for these new (sub)-Actions, although there is still Boson/Fermion mass degeneracy in a (sub)-Action, so long as there is still a Boson present.  Notably, there is even a SUSY (sub)-Action where the physical Scalars are not present, so that  the (sub)-Action contains physical Fermions only.  In this theory the degeneracy of Bosonic and Fermionic masses is obviously not present, and this happens without the spontaneous or explicit breaking of SUSY.  No such `Exchange Transformation' has been found for SUSY gauge Actions.

\refstepcounter{orange}\la{intro}
{\bf \theorange}.\;
{\bf Introduction}:
In spite of a large amount of work, there is still much that is unknown about SUSY
\ci{west,superspace,WB,ferrarabook,KBbook,Weinberg3}.
Even its representation theory is filled with unanswered questions \ci{Gatesfundremains}. 
Much recent work has concerned itself with  geometric and duality issues in supersymmetric theories.  Some of this is associated with   
Branes, M theory and AdS CFT \ci{texts,Freedman:2012zz,newwest,becker}.
 A large body of SUSY work has, understandably,  focussed on phenomenology and experimental signals, while leaving the problems of the origin of the  spectrum of mass splitting for future work \ci{xerxes,haber,susymeets}. The progress reported in this paper is the direct result of a study of the algebraic problem of the local BRST cohomology of the \CM\ \ci{one,two,three}, which examined the solution to the tachyon problem.  Previous work, troubled by the tachyon problem,  was in \cite{johnsusy,Dixonholes, DixonChiralcohom, Dixon:1993jt, DixonRahm, Dixonjumps} using some of the methods in    \ci{Dixonspec, Mcleary}.
All of that work was based on the  \PB\ formulation of symmetries \ci{poissonbrak,Becchi:1975nq,zinnbook,taylor,Zinnarticle,Weinberg2}.

From the beginning of research into 
SUSY, it has been noticed by many authors that 
 SUSY seems to
 hint at solutions
to various problems.  But these hints then turn to disappointment, because the effort to remove the mass
degeneracy of the supermultiplets, using spontaneous SUSY breaking, tends to spoil the nice properties of the theory
\ci{weinbergcosmo}. This has led some authors to wonder whether the mass degeneracy of SUSY can be removed, even when SUSY itself 
is not really spontaneously (or explicitly) broken at all \ci{wittenreally}.  

 The theory presented here shows how the mass degeneracy can be removed for Chiral Multiplets, without spontaneous or explicit breaking of SUSY.  The result is a sort of compromise between the usual SUSY theories that have a conserved SUSY charge, and theories which have no SUSY at all.  We use an Exchange Transformation\footnote{This is like a canonical transformation--it ensures that the new Master Equation yields zero with the new Action.  See \pgh\ \ref{aboutCTRs} below.} to change the original normal SUSY theory to a theory with a non-conserved, but still very relevant,  
SUSY charge. The new theory satisfies a Master Equation that is very similar to the Master Equation of the parent SUSY theory.
 The method here does not apply to SUSY Gauge theory so as to remove its SUSY charge.  But the presence of SUSY Gauge theory is not a problem for the method.

\refstepcounter{orange}\la{intro2}
{\bf \theorange}.\;
{\bf \HCM s and \UCM s:}  For each \CM\ in an Action, there is a choice to be made.  One can simply leave it as a \CM, or one can use  an \CTR\ to transform it to a new kind of Multiplet, with a new Action and new physics.  This \CTR\ transforms some or all of  the Scalar Field $A$ to a  Zinn  Source\footnote{\la{aboutmaster} The original formulation of the Master Equation as an `antibracket' was by Zinn-Justin \ci{poissonbrak,Becchi:1975nq,zinnbook,taylor,Zinnarticle,Weinberg2}. 
The Sources in the \PB\ are sometimes called `Antifields',  following \ci{BV}. The author thinks that this term is   misleading and confusing. The so-called `Antifields' are very definitely not Fields--they are Sources. The term `anti' is also confusing, because  the term `Antifield' would naturally means the `Field that creates Antiparticles'.  So here these Sources are called Zinn Sources.   The Zinn Sources  do not get integrated in the Feynman path integral, whereas Fields do get integrated, and so, of course, do their \CC s, the Antifields. We use the term `Zinn Action' to denote the part of the Action that is at least linear in  Zinn Sources. It is unphysical, but useful to keep track of the symmetry.}
 $J$ and the Zinn Source $\G$  to an Antighost Field $\h$.
The Zinn Source $\G$ is the Source for variations of the Scalar Field $A$ (in the old Action). The  Zinn Source $J$ is the Source  
 for variations of the Antighost Field $\h$ (in the new Action).

 The  \CTR s of the \PB\  that we will introduce in this paper will be said to give rise to a  {\bf \HCM}, when one Scalar remains, and to an {\bf \UCM}, when no Scalars remain.  This decision can be made for each \CM\ in the theory, separately. Since the two Scalars in a Chiral Multiplet are not equivalent to each other (they differ in parity for example), there are really three different \CTR s possible for each \CM. The \CTR\ is, in terms of its construction,  a sort of Canonical Transformation, except that it takes one theory to a different theory for the \PB\ case.  This is explained in Section \ref{aboutCTRs} below.

The \HCM s are useful for  the spontaneous breaking of Gauge symmetry.  However, some of the mass degeneracy survives for the Gauge/Higgs sector, as will be shown in \ci{five}. The \UCM s are useful to make an Action with no Scalars, and we will use them  for the SSM in \ci{five}, to eliminate all the Squarks and Sleptons, leaving  no mass degeneracy  in that Matter sector.
Clearly this is a  form of `SUSY Charge Suppression', which eliminates the SUSY charge in a particular sector, while preserving the Master Equation, which has a strong influence from  
SUSY.

A pleasant feature is that the \CTR\ also allows us to construct interactions for these  new `Suppressed SUSY' theories. These follow directly from the initial Action, which is made entirely from Gauge and \CM s (and ultimately \SG). The \CTR\  also allows one to show, in a formal way, that, in perturbation theory, one can  construct new Quantum Field theories from the new theories.  These  preserve SUSY with the new \PB, as is shown in \pgh s \ref{derivenewbrstforUCM} and  \ref{derivefieldtransforUCM} for the \UCM\ case. A similar derivation can be done for the \HCM.

\refstepcounter{orange}\la{summ}
{\bf \theorange}.\;
{\bf This paper
is organized as follows:} We introduced some names for the new theories in  \pgh\ \ref{intro2}.  Next, in \pgh\ \ref{objections}, we will pose and answer some questions that are bound to arise in the mind of a reader who is familiar with SUSY. 
Then the way that these new Multiplets arose from the study of the BRST cohomology 
is described in  \pgh\   \ref{Chiralcohom}.
The Chiral Multiplet and the integration of its auxiliary field is in \pgh s \ref{Chiraltheory} to \ref{theoremaboutaux}.  
The Majorana version of the \HCM\ is explained starting with \pgh\  \ref{MajICDSS}
to   \ref{MajHClastnowchiral}.
 \pgh\   \ref{aboutCTRs} dicusses how and why these \CTR s are special.
Then in \pgh\ \ref{backwards}, we show how to add a mass term to the \HCM.  One starts by adding a mass term to the \CM, which was already done in \pgh\  \ref{Chiraltheory} .  Then   the \CTR\ is used to convert that Action to the \HCM\ Action.  The result is interesting, because it is easy to see how the SUSY Charge is modified, as is explained in
\pgh\ \ref{backwardsconfusion}. Interactions in the \HCM\ are discussed in \pgh\ \ref{withinteraction}.  
Then we introduce the \UCM\ in \pgh\ \ref{UCM} . 
Again we introduce a mass term by starting with the massive \CM, and ending with the massive \UCM\ in 
\pgh\ \ref{massforUCM}. This case is simpler than the \HCM\ case, because there are only fermions left in the \UCM.
 Next the well known derivation of the \PB\ for the \CM\ is recalled in \pgh\ \ref{derivenewbrstforCM}.  This is  followed by the parallel derivation for the \UCM\ in \pgh\  \ref{derivenewbrstforUCM}.  This latter derivation depends crucially on the  invariance of the \PB\ under   the \CTR,  as shown in \pgh\  \ref{derivefieldtransforUCM}.  
Some remarks about the Hamiltonian and the Cohomology are made in \pgh\ \ref{hamsection}. The Conclusion summarizes the results in {section} \ref{conclusion2}.
This is  followed by a summary of  the Dirac version of the \HCM\  in the Appendix, which occupies   \pgh s  \ref{diraccase} to \ref{MCPB}.

\refstepcounter{orange}\la{objections}
{\bf \theorange}.\;
 {\bf Surely the Chiral Multiplet is irreducible, so how can it be reduced? }
{\bf And surely the SUSY algebra implies Degenerate Super-Multiplets \ci{algebrarefs}, so how can there be field theories which do not have them unless  SUSY itself is  spontaneously or explicitly broken?
And if the SUSY charge is not conserved, how can there be any SUSY left at all?}

{\bf Answer:} 
\bitem
\item
To see in detail how the theory gets around these very reasonable objections, the reader (if he or she is impatient) can look at \pgh\ \ref{massforUCM} 
for the \UCM\ case, which is rather simple.  The \HCM\ case, which is a little more complicated, 
is explained in \pgh s
\ref{backwards},  \ref{massandint} and 
\ref{backwardsconfusion}. 
\item
The result of the \CTR\ is not a new representation of SUSY.  It is  a representation of the SUSY algebra in terms of Fields and Zinn Sources, which is physically not the same at all as a representation in terms of Fields alone.  Fields are physically realizable, and Zinn Sources are not.  When the SUSY algebra spreads to the Zinn Sources, it is no longer realized in the same way on the physical states.
\item
If  \HCM s or \UCM s are present, then an attempt to construct a 
Noether SUSY Current yields an expression which is dependent on the Zinn Sources.   So the SUSY Charge 
cannot be  constructed\footnote{We recall the Noether analysis: an invariance of the Action leads to a current ${\cal J}_{\m}$ that has zero divergence $\pa^{\m} {\cal J}_{\m}=0$.  Then the integral 
$Q= \int d^3 x {\cal J}_{0}$ satisfies $\fr{d}{d t}Q=0$, assuming that the surface integral $\int dS^i {\cal J}_{i}$ goes to zero at spatial infinity. The demonstration of this requires use of the Equations of Motion of the Fields.  This is explained in most texts on Quantum Field theory.}, because the Zinn Sources do not satisfy equations of   motion. However, as we shall see, this  `Suppressed SUSY Charge theory' 
still  has Bose/Fermi mass degeneracy in the 
 \HCM\ case.  But it cannot do so for the \UCM, because in that case there are no bosons left in the multiplet.
\item
The relation of this `Suppressed SUSY Charge' mechanism to the BRST cohomology is rather obscure.  
This is true even though 
the examination of BRST cohomology is how it was found, as will be explained below in \pgh\ \ref{Chiralcohom}.
One could say that the SUSY chiral algebra is still present, although the conserved SUSY charge is not.
This puzzling topic, and the question of the Hamiltonian, are discussed a little more in \pgh\ \ref{hamsection}.
\eitem

\refstepcounter{orange}\la{Chiralcohom}
{\bf \theorange}.\;
{\bf BRST Cohomology of the Chiral Multiplet}:
 The BRST cohomology of  the \CM\ is immense, but  much of it  has unsaturated Lorentz Spinor indices  \ci{johnsusy}.  By and large, this  has been taken to mean that  `{\em this cohomology  cannot appear in a Lorentz invariant Action and so it is  of no interest}'.  {\bf But this pessimistic view is not valid}.  The cohomology is very important, because it can only be relevant if one  couples the unsaturated indices to something new, so that the cohomology 
can appear in an Action. {\em The point is that this opens up a new view on SUSY. } The simplest such new object is evidently a \cdss\ci{one}.  But then the question is whether such an object makes sense by itself, and the answer has been discouraging for a long time.   The most obvious Action for the \cdss\ has higher derivatives and also tachyons.  But recently some progress was made on this problem \ci{one,two,three}. A tachyon free Action for the \ICDSS\ was found and used in a free theory\footnote{That progress was not very exciting however, because it dealt only with free theories.  Moreover, efforts to make those theories interacting have been fraught with obstructions of the kind mentioned in \ci{two}.  However this changes if one integrates the auxiliary $W$, as will be seen below.}.
These \CTR s were found while trying to make that theory an interacting one. 
In fact they go beyond the results from the cohomology.   We will return to the Majorana version of the \ICDSS\ in \pgh\ \ref{MajICDSS}.  The Dirac version of the \ICDSS\  is in the Appendix starting with \pgh\ \ref{diraccase}.

\refstepcounter{orange} \la{Chiraltheory}
{\bf \theorange}.\;{\bf The Chiral Scalar Superfield}: We start with the well-known \CM\ theory.  It has the total  Action:
\be
{\cal A}_{\rm Chiral\;Total}
={\cal A}_{\rm Chiral\;Kinetic}
+ 
{\cal A}_{\rm Chiral\;Mass\;and\;Interaction}
+
{\cal A}_{\rm Chiral\; Zinn}
\ee
  where the free massless
kinetic Action is:
\be
{\cal A}_{\rm Chiral\;Kinetic}
= \int d^4 x \;\lt \{
 F
\ov F
-
\y_{\a  }     \pa^{\a \dot \beta }   
{\ov \y}_{ \dot \beta}
+
\fr{1}{2} 
\pa_{ \a \dot \beta }   A    \pa^{\a \dot \beta  } {\ov  A} 
\rt \}
\la{Chiralkinetic}
\ee
and mass and interaction terms look like\footnote{Here $+*$ means `add the \CC\ of the previous terms'.} 
\be
{\cal A}_{\rm Chiral\;Mass\;and\;Interaction}
= \int d^4 x \;\lt \{
m_1 A  F -
\fr{1}{2} 
 m_1 \y^{\a }   
\y_{\a} 
+ g_1
 A^2  F -
g_1  A \y^{\a }   
\y_{\a} 
+ *
\rt \}
\la{Chiralmassint}
\ee

and the Zinn Action
 is:
\be
{\cal A}_{\rm Chiral\; Zinn}
=
\lt \{
\G  
\lt (
C^{ \a}  \y_{   \a}
+ \xp A \rt )
+ Y^{\a} \lt (
\oC^{\dot\d}  \pa_{\a \dot \d}
 A 
+
C_{\a}
 F
+ \xp \y_{\a} \rt )
\ebp+ \Lam 
\lt ( \oC^{\dot \b} 
    \pa_{\a \dot \b} 
 \y^{ \a}
+ \xp F \rt )
+*
\rt \}
+
\cA_{\rm SUSY}
\la{fullChiral}
\ee
where
\be
\cA_{\rm SUSY}
= - C^{\a} \oC^{\dot \a} h_{\a \dot \a}
\ee
Here $A$ is a complex Scalar Field, $\y_{\a}$ is a two-component Weyl Spinor Field, and  $F$ is a complex auxiliary Scalar Field.  The Zinn Action  has the form 
$\ic \lt \{ \G \d A + Y \d \y + \Lam \d F + *\rt \}$, which is a sum of Zinn Sources
  coupled to the SUSY variations of the Fields.

The SUSY invariance can be summarized by the fact that the above Action ${\cal A}_{\rm Total}
$ yields zero for the \PB:
\be
\cP_{\rm Chiral}
 [\cA]=
\int d^4 x \lt \{
\fr{\d \cA}{ \d A}
\fr{\d \cA}{ \d \G}
+
\fr{\d \cA}{ \d  \y_{ \a}}
\fr{\d \cA}{ \d Y^{ \a}}
+\fr{\d \cA}{ \d F}
\fr{\d \cA}{ \d \Lam} +*
\rt \}+\fr{\pa \cA}{ \pa h_{\a\dot \a} }
\fr{\pa \cA}{ \pa \x^{\a\dot \a} }
\ee
We define
\be
\x \cdot \pa
\equiv
\x^{\a \dot \a}
\pa_{\a \dot \a}
\ee
The following terms\footnote{\la{coupletoSG} One cannot implement the BRST method unless one closes the algebra.  But there is a genuine problem here.  Although there are identities which arise from this \PB\ that are true in perturbation theory for this  rigid SUSY theory,  those that involve $\x$ are not true, because $\x$ is not integrated in the Feynman path integral, although it is more like a Field than a Zinn Source. This problem can be resolved by embedding the rigid theory in \SG.} must  always be added to every  \PB\ of SUSY theories:
\be
\cP_{\rm SUSY}[\cA]
=
\fr{\pa \cA}{ \pa h_{\a\dot \a} }
\fr{\pa \cA}{ \pa \x^{\a\dot \a} }
= -
C_{\a} \oC_{\dot \a} 
\fr{\pa \cA}{ \pa \x^{\a\dot \a} }
\la{susypb}
\ee
These terms take account of the fact that the SUSY algebra closes onto a translation. The BRST ghost of that translation for rigid SUSY is the constant anticommuting vector $\x^{\a\dot \a}$.  Equation (\ref{susypb}) summarizes the fact that two SUSY transformations with the SUSY parameter $C_{\a}$ act like a spacetime translation.

\refstepcounter{orange} \la{Chiraltheory2}
{\bf \theorange}.\;{\bf The Chiral Scalar Theory after Integration of the Auxiliary $F$} 
\la{baskcther}
After integrating the $F$ auxiliary Field in  
 \pgh\  \ref{Chiraltheory}, and dropping the Source $\Lam$ for its variation, one gets\footnote{Insert the Action into a Feynman path integral with Sources for the Fields, as in \pgh\ \ref{derivenewbrstforCM} below, and derive the \PB\ in the usual way.  Then complete the quadratic in $F$ and $\oF$ and perform the same exercise, after dropping the Source $\Lam$.  Then shift the $F$ and  integrate it, which just leaves a number.  This leaves the terms and Zinn Action shown.}.
\be
{\cal A}_{\rm Chiral\;F\;Int }
= \int d^4 x \;\lt \{
-
\y_{\a  }     \pa^{\a \dot \beta }   
{\ov \y}_{ \dot \beta}
+
\fr{1}{2} 
\pa_{ \a \dot \beta }   A    \pa^{\a \dot \beta  } {\ov  A} 
+
\G  
 \lt (C^{ \a}  \y_{   \a}
+ \xp A \rt )
\ebp+ Y^{\a}
\lt ( \oC^{\dot\a}  \pa_{\a \dot \a}
 A  + \xp \y_{\a} \rt )
+
\oG  
\lt (
\oC^{\dot \a}  \oy_{\dot   \a}
+ \xp \A\rt )
+ \oY^{\dot \a}
\lt (
C^{\a}  \pa_{\a \dot \a}
 \A 
+ \xp \oy_{\dot \a}
\rt )
\ebp
 -
\fr{1}{2} 
 m_1 \y^{\a }   
\y_{\a} 
 -
g_1  A \y^{\a }   
\y_{\a} 
 -
\fr{1}{2} 
 \om_1 
\oy^{\dot\a }   
\oy_{\dot \a} 
 -
\og_1  \A 
\oy^{\dot\a }   
\oy_{\dot \a} 
\ebp-
\lt (   \om_1 \A + \og_1 \A^2 + \oY^{\dot\a} \oC_{\dot \a} \rt )
\lt (   m_1 A + g_1 A^2 + Y^{\a} C_{\a} \rt )
\rt \}
\ee
which yields zero for the smaller \PB:
\be
\cP_{\rm Chiral\; F\;Int }[\cA]=
\int d^4 x \lt \{
\fr{\d \cA}{ \d \A}
\fr{\d \cA}{ \d \ov\G}
+
\fr{\d \cA}{ \d  \oy_{\dot  \a}}
\fr{\d \cA}{ \d \oY^{\dot \a}}
+
\fr{\d \cA}{ \d A}
\fr{\d \cA}{ \d \G}
+
\fr{\d \cA}{ \d  \y_{  \a}}
\fr{\d \cA}{ \d Y^{ \a}}
\rt \} 
+\fr{\pa \cA}{ \pa h_{\a\dot \a} }
\fr{\pa \cA}{ \pa \x^{\a\dot \a} }\la{Chiralzinn}
\ee
This is the \PB\ 
 for a  \CM\ where the auxiliary $F$
has been integrated out.  We will see this form again below.  The Zinn Sources $Y$ appear quadratically, and they keep the invariance intact. Now we shall put this derivation in a theorem that we will often need:

\refstepcounter{orange}\la{theoremaboutaux}
{\bf \theorange}.\;
{\bf Theorem about \AF s and \PB s}:
The technique in the above example is frequently used in this paper, and it is worth making it into a theorem. We will use this theorem repeatedly in this paper and the paper \ci{five}.

{\bf Theorem 1}: Given an Action $\cA$ that satisfies a given \PB\ $\cP$,  then 
\ben
\item
Suppose that  there is a Field F
in that Action\footnote{Here we assume that $F$ is real.  In \pgh\  \ref{baskcther}, F was complex.} which  has an algebraically invertible quadratic term, and a linear term\footnote{Auxiliary Fields generally satisfy this condition.} in $F$, so that the total Action has the form:
\be
\cA= \ic \lt \{ m_{ij} F^i F^j + F^i G_i   +\Lam_i \d F^i \etc
\rt \} 
+\cA_{\rm Other \;Terms}
\la{termswithF}
\ee
and that the \PB\ has the form
\be
\cP[\cA] =\ic  \lt (
\fr{\d \cA}{ \d \Lam_i }
\fr{\d \cA}{ \d F^i }\rt ) + \cP_{\rm Other \;Terms}[\cA]
\la{PBFRT}
\ee
\item
Then we can integrate out the Field $F$ and get a new Action and a new \PB\  as follows:
\ben
\item
Remove the Zinn Source term $\ic \Lam_i \d F^i$ from the Action (\ref{termswithF}), and 
\item
Remove the related term $
\fr{\d \cA}{ \d \Lam_i }
\fr{\d \cA}{ \d F^i }$
from the \PB\ (\ref{PBFRT}),
\item
The new Action is 
\be
\cA_{\rm New}=\ic \fr{-1}{4}\lt \{ (m^{-1})^{ij} G_i G_j   \rt \}
+\cA_{\rm Other \;Terms}
\ee
\item
The new Action yields zero for the new \PB, which reduces to
\be
\cP_{\rm New}[\cA] = \cP_{\rm Other \;Terms}[\cA]
\ee

\een
\een
The proof is simple. Write the relevant terms in the Action (\ref{termswithF})
in the form
\be
\cA=\ic \lt \{ 
m_{ij} \lt ( F^i + \lt (\fr{m^{-1}}{2} G\rt )^i \rt )\lt ( F^j + \lt(\fr{m^{-1}}{2} G\rt )^j \rt )
- m_{ij}   \lt (\fr{m^{-1}}{2} G\rt )^i   \lt(\fr{m^{-1}}{2} G\rt )^j 
\rt \}
\ee
Then shift and integrate the Field $F^i$, after placing it in a Feynman path integral as in \pgh s \ref{derivenewbrstforCM} and \ref{derivenewbrstforUCM} below.  This yields an irrelevant constant plus the second term. It is important that the $G^i$ does not contain $F^i$.  But it can contain anything else including Zinn Sources.

 The usual demonstration that the  \PB\ yields zero, as in \pgh s \ref{derivenewbrstforCM} and \ref{derivenewbrstforUCM} below, goes through when this has been done, and no Zinn Source for the variation of $F^i$ is needed because $F^i$ is gone from the theory.  This theorem is important because our \CTR s here map  Actions which have had their auxiliaries integrated, as we shall see.

\refstepcounter{orange}\la{MajICDSS} 
{\bf \theorange}.\;
{\bf The Majorana  \ICDSM:} 
Next we consider  the simplest example of a Majorana \ICDSM.  It is the  Majorana version of the Dirac type theory\footnote{This could be derived in  exactly the same way as in \ci{one,two,three}, using BRST recycling, starting with the U(1) Gauge theory in this Majorana case. We start with this simplest case but the other case is also needed and we return to it below in \pgh\ \ref{diraccase}} that was discussed in \ci{one,two,three}. We will use the notation of  \ci{one}. The Action is quite simple:
\be
{\cal A}_{\rm MI}
=
{\cal A}_{\rm MI\;Kinetic}
+ 
\cA_{\rm MI\; Zinn}
\la{MHCAction}
\ee
where
\be
{\cal A}_{\rm MI\;Kinetic}
=  \int d^4 x \; \lt \{
-\f^{\dot \a}
\pa_{\a\dot\a}
\ov\f^{ \a}
-\fr{1}{2}
 W_{\a\dot \a}
W^{\a\dot \a}
+
\fr{1}{2}G \Box G 
-
\fr{1}{\sqrt{2}} \eta  \lt (
 \f^{\dot \d}
\oC_{\dot \d}
+
\ov\f^{\d}
C_{ \d}
\rt )
\rt \}
\la{MHCkinetic}
\ee
In the above, $G$ is a real Scalar Field, $\f^{\dot \a}$ is a two component complex Weyl
Spinor, $W_{\m}$  is a real vector Field\footnote{$W_{\a \dot \a} = W_{\m}\s^{\m}_{\a \dot \a}$}, and it turns out to be auxiliary (no kinetic term), $\h$ is a real Grassmann odd Antighost Scalar Field and $C_{\a}$ is again the Grassmann even space-time constant Weyl Spinor ghost Field.  The Zinn Action that we need is:

\[
\cA_{\rm MI\; Zinn}
=
\int d^4 x
 Z^{\dot \a}
\lt (
-i\fr{1}{\sqrt{2}} \pa_{\a \dot \a} G
C^{\a} 
-
W_{\a \dot \a} C^{\a}
+
\x \cdot \pa\;\f_{ \dot \a} 
\rt ) 
+
\ov Z^{ \a}
\lt (
i\fr{1}{\sqrt{2}} \pa_{\a \dot \a} G
\oC^{\dot\a}
-
W_{\a \dot \a} \oC^{\dot\a}
+
\x \cdot \pa\;\ov \f_{  \a} 
\rt ) 
\]
\[ 
+
\S^{\a\dot \a}
\lt (
 \sqrt{2} \eta   \oC_{\dot \a}
C_{\a}
-\fr{1}{2}
\pa_{\a}^{\dot \g} \f_{  \dot \g} 
 \oC_{\dot \a}
-\fr{1}{2}
\pa_{\a}^{\dot \g} \f_{  \dot \a} 
 \oC_{\dot \g}
-\fr{1}{2}\pa_{\dot\a}^{  \g} 
\ov \f_{    \g} 
 C_{  \a}
-\fr{1}{2}\pa_{\dot\a}^{  \g} 
\ov \f_{    \a} 
 C_{  \g}
+ \x \cdot \pa\;
 W_{\a\dot \a}\rt )
\] 
\be 
+
   \U
\lt ( 
- \fr{i}{ \sqrt{2}}\ov\f_{ \b}
C^{ \b}
+\fr{i}{ \sqrt{2}}
 \f_{\dot \b}
\oC^{\dot \b}
+\x \cdot \pa G
\rt ) 
+
 J
\lt (
 \fr{1}{\sqrt{2}}
\pa_{\g  \dot \d} 
 W^{\g  \dot \d}  
  +   \x \cdot \pa  \eta \rt )
+\cA_{\rm SUSY}
\ee

Any of the three Actions in (\ref{MHCAction})
 yields zero when inserted into the following \PB\ for SUSY:
\be
\cP_{\rm MI}[\cA]
=
\int d^4 x \lt \{
\fr{\delta \cA}{ \delta Z^{\dot \a}}
\fr{\delta \cA}{ \delta \f_{\dot \a}}
+
\fr{\delta \cA}{ \delta \oZ^{  \a}}
\fr{\delta \cA}{ \delta \ov \f_{ \a}}
+
\fr{\delta \cA}{ \delta \S^{\a\dot \a}}
\fr{\delta \cA}{ \delta W_{\a\dot \a}}
+
\fr{\delta \cA}{ \delta  \U}
\fr{\delta \cA}{ \delta G }
+\fr{\delta \cA}{ \delta  J}
\fr{\delta \cA}{ \delta \eta }
\rt \}+\fr{\pa \cA}{ \pa h_{\a\dot \a} }
\fr{\pa \cA}{ \pa \x^{\a\dot \a} }
\ee

\refstepcounter{orange}\la{ICDSM}
{\bf \theorange}.\;
{\bf Integrate the auxiliary out of the \ICDSM}:  After \ci{one,two,three} were written, the main problem for the new version of the \ICDSS\ was whether it could be put into an interacting Action. That seemed very difficult at first.  However, it turns out that 
interactions can be generated easily if we first integrate  the auxiliary vector Field $W_{\a \dot \b}$ out of  the Action for the Majorana \ICDSS\ in \pgh\ \ref{MajICDSS}. 
We will give the result of that a new name:   the  \HCM\ Action.
We use that name for the \HCM\ because it has half the Scalar degrees of freedom that a \CM\ has.

 Once the integration of $W$  is done, it is fairly easy to recognize that the resulting \HCM\ Action  is really  the result of an \CTR\ acting on a \CM\ that has had its auxiliary Field $F$ integrated.   This is the \CTR\ that we will be using.  Because it will turn out that this theory can be obtained from an \CTR\ acting on a \CM, we can  couple the \CM\ using known methods, and then using the inverse \CTR, we can deduce the interactions of the \HCM.

\refstepcounter{orange}\la{MHCM}
{\bf \theorange}.\;{\bf The \MHCM:}
Drop the $\S$ terms in the action in Equation (\ref{MHCAction})  in \pgh\ \ref{MajICDSS}, 
and integrate $W$ out of the Action\footnote{Insert the Action into a Feynman path integral with Sources for the Fields, and derive the \PB\ in the usual way.  Then complete the quadratic in $W$ and perform the same exercize, while leaving its Source out, and shifting the W to integrate it.  This leaves the terms and Zinn Action shown. This is an application of the theorem in \pgh\ \ref{theoremaboutaux}.}.  This yields a closely related Action, which we will dignify by a new name, the \MHCM:

\[
{\cal A}_{\rm MHC}
=  \int d^4 x \; \lt \{
-\f^{\dot \a}
\pa_{\a\dot\a}
\ov\f^{ \a}
+
\fr{1}{2}G \Box G 
-
\fr{1}{\sqrt{2}} \eta  \lt (
 \f^{\dot \d}
\oC_{\dot \d}
+
\ov\f^{\d}
C_{ \d}
\rt )
\rt \}
\]

\[
+
\int d^4 x
\lt \{
 Z^{\dot \a}
\lt (
-i\fr{1}{\sqrt{2}} \pa_{\a \dot \a} G
C^{\a} 
+
\x \cdot \pa\;\f_{ \dot \a} 
\rt ) 
\rt.
+
\ov Z^{ \a}
\lt (
i\fr{1}{\sqrt{2}} \pa_{\a \dot \a} G
\oC^{\dot\a}
+
\x \cdot \pa\;\ov \f_{  \a} 
\rt ) 
\]
\[ 
\lt.+
   \U
\lt ( 
- \fr{i}{ \sqrt{2}}\ov\f_{ \b}
C^{ \b}
+\fr{i}{ \sqrt{2}}
 \f_{\dot \b}
\oC^{\dot \b}
+\x \cdot \pa G
\rt ) + J \xp \h 
\rt \}
\]
\be
+ \fr{1}{2}
 \int d^4 x \; 
\lt \{
 Z^{\dot \a}
 C^{\a}
+
 \ov Z^{  \a}
 \oC^{\dot \a}
+
 \fr{1}{\sqrt{2}}
\pa^{\a  \dot \a} 
 J
\rt \}
\lt \{
 Z_{\dot \a}
 C_{\a}
+
 \ov Z_{  \a}
 \oC_{\dot \a}
+
 \fr{1}{\sqrt{2}}
\pa_{\a  \dot \a} 
 J
\rt \}
+\cA_{\rm SUSY}
\ee
which we can rearrange to
\[
{\cal A}_{\rm MHC}
=  \int d^4 x \; \lt \{
-\f^{\dot \a}
\pa_{\a\dot\a}
\ov\f^{ \a}
+
\fr{1}{2}G \Box G 
+
\fr{1}{2}J \Box J -
\fr{1}{\sqrt{2}} \eta  \lt (
 \f^{\dot \d}
\oC_{\dot \d}
+
\ov\f^{\d}
C_{ \d}
\rt )
\rt.
\]\[
\lt.+
   \U
\lt ( 
- \fr{i}{ \sqrt{2}}\ov\f_{ \b}
C^{ \b}
+\fr{i}{ \sqrt{2}}
 \f_{\dot \b}
\oC^{\dot \b}
+\x \cdot \pa G
\rt ) 
\rt \}
\]

\[
+
\int d^4 x
 Z^{\dot \a}
\lt (
 C^{\a}
 \fr{1}{\sqrt{2}}
\pa_{\a  \dot \a} 
 J
-i\fr{1}{\sqrt{2}} \pa_{\a \dot \a} G
C^{\a} 
+
\x \cdot \pa\;\f_{ \dot \a} 
\rt ) 
\]
\[
+
\ov Z^{ \a}
\lt (
 \oC^{\dot \a}
 \fr{1}{\sqrt{2}}
\pa_{\a  \dot \a} 
 J
+i\fr{1}{\sqrt{2}} \pa_{\a \dot \a} G
\oC^{\dot\a}
+
\x \cdot \pa\;\ov \f_{  \a} 
\rt ) 
\]
\be
+
 \int d^4 x \; 
\lt \{
+
 Z^{\dot \a}
 C^{\a}
 \ov Z_{  \a}
 \oC_{\dot \a}
+ J \xp \h \rt \}
+\cA_{\rm SUSY}
\la{bertterfrm}
\ee

Now this yields zero for:
\be
\cP_{\rm MHC}[\cA]
=
\int d^4 x \lt \{
\fr{\delta \cA}{ \delta Z^{\dot \a}}
\fr{\delta \cA}{ \delta \f_{\dot \a}}
+
\fr{\delta \cA}{ \delta \oZ^{  \a}}
\fr{\delta \cA}{ \delta \ov \f_{ \a}}
+
\fr{\delta \cA}{ \delta  \U}
\fr{\delta \cA}{ \delta G }
+\fr{\delta \cA}{ \delta  J}
\fr{\delta \cA}{ \delta \eta }
\rt \}+\fr{\pa \cA}{ \pa h_{\a\dot \a} }
\fr{\pa \cA}{ \pa \x^{\a\dot \a} }
\la{mhcmasterdf}
\ee

\stepcounter{orange} \la{majremark2}
{\bf \theorange}. \; {\bf Remarkable Symmetries of  the Action  $\cA_{\rm MHC}
$ :}
The above Action $\cA_{\rm MHC}
$ in Equation (\ref{bertterfrm})
 has a remarkable symmetry which was not obvious before we integrated the auxiliary $W$. The Field $\h$ and the Source $\U$ appear in similar ways. 
  The term 
$ \eta  \lt (
 \fr{1}{ \sqrt{2} } \ov\f_{  \d}
C^{\d}
+\fr{1}{ \sqrt{2} }  
 \f_{ \dot\d}
\oC^{\dot \d}
\rt )
$ looks like a Zinn Source coupled to a variation, except that $\h$ is a Field.
 The Field $G$ and the Source $J$ also appear in similar ways. The term $\fr{1}{ 2 }
 J \Box
J$ looks like a kinetic term for $J$, except that $J$ is a Source.

\refstepcounter{orange}\label{hccantran} 
{\bf \theorange}.\;
{\bf New Variables for the \MHCM}
 This symmetry can be exploited with a  Generating  Functional   for an \CTR\ of the Action and \PB.  The new Action  will yield zero for  the new \PB. Our new Action will have `a new complex Field' $({S}, {\oS})$ and `a new complex Zinn Source'
$(\G,\ov \G)$.  These will replace
the `old real Fields'  $(\h ,G)$
 and the `old real Zinn Sources' $(J, \U)$.
We will choose a Generating  Functional  of the new
Zinn Sources $(\G,\ov \G)$ and the old Field $G$ and the old Zinn Source $J$.

\be
\cG_{\rm  MHC} = 
\int d^4 x \lt \{
 \fr{1}{\sqrt{2}}\G (J -iG )
+
 \fr{1}{\sqrt{2}}\oG (J +iG )
 \rt \}
\la{majoranagenfunc}
\ee
and the \CTR s are:
\be
S=\fr{\d \cG}{\d \G}=
 \fr{1}{\sqrt{2}}(J-iG)
;\;\oS=\fr{\d \cG}{\d \ov\G}=
 \fr{1}{\sqrt{2}}(J+iG)
;\; \eb
\h=\fr{\d \cG}{\d J}
=
 \fr{1}{\sqrt{2}}\lt (
\G+ \oG
\rt )
;\;\U=\fr{\d \cG}{\d G}=
 \fr{1}{\sqrt{2}}(-i\G+ i \oG)
\ee
These have the following inverses:
\be
G= 
 \fr{1}{\sqrt{2}}(i S- i \oS)  
;\;J= 
 \fr{1}{\sqrt{2}}(S  +  \oS )
;\;\G= 
 \fr{1}{\sqrt{2}}(\h + i \U) 
;\;\ov\G= 
 \fr{1}{\sqrt{2}}(\h - i \U ) 
\ee

\refstepcounter{orange}\la{MajHClastnowchiral}
{\bf \theorange}.\;{\bf New Action after \CTR: It looks like the Chiral Action} 
The new Action expressed in terms  of the new variables is:
\[
{\cal A}_{\rm CM}
\la{Majkin}
=  \int d^4 x \; \lt \{
-\f^{ \dot \a}
\pa_{\a\dot\a}
\ov\f^{\a}
+
{S} \Box {\oS} 
+
\oG\lt (
\ov\f_{\d}
C^{ \d}
+ \xp {\oS}\rt )
+
\G
\lt (\f_{\dot\d}\oC^{\dot \d}
+ \xp {S}\rt )
\rt. \]
\be
\lt.+
  Z^{\dot \a} 
\lt (
  \pa_{\a \dot \a} S
C^{\a}
+ \xp \f_{\dot\a}\rt )
 + \ov Z^{ \a} 
\lt (
\pa_{\a \dot \a} \oS
\oC^{\dot\a}  
+ \xp
\ov \f_{\a}  
\rt ) 
- 
   Z_{\dot \a} C^{\a}
\ov Z_{ \a} \oC^{\dot\a}  
\rt \}
\la{MMM}
\ee
The new \PB\ is:
\be
\cP_{\rm CM}[\cA]
=
\int d^4 x \lt \{
\fr{\delta \cA}{ \delta Z^{\dot \a}}
\fr{\delta \cA}{ \delta \f_{\dot \a}}
+
\fr{\delta \cA}{ \delta  \G}
\fr{\delta \cA}{ \delta {S}}
+*
\rt \}
+\fr{\pa \cA}{ \pa h_{\a\dot \a} }
\fr{\pa \cA}{ \pa \x^{\a\dot \a} }\ee
The invariance of the new Action  $\cA_{\rm CM}$ is expressed by:
\be
\cP_{\rm CM}[ \cA_{\rm CM}]=0
\ee
Note that the expressions in this Section are  identical to the results for the \CM\ in \pgh\ \ref{Chiraltheory2}
 above, if one changes the names of the Fields and Zinn Sources, and sets $m_1=g_1=0$ in \pgh\ \ref{Chiraltheory2}.
Here is the mapping from \pgh\ \ref{Chiraltheory2} to this \pgh\ \ref{MajHClastnowchiral}.
\be
A \ra \oS;
\G \ra \oG;
Y^{ \a} \ra \oZ^{ \a};
\y_{ \a}
\ra \ov \f_{\a}
\ee
and their \CC s.

\refstepcounter{orange}\label{aboutCTRs} 
{\bf \theorange}.\;
{\bf Poisson Brackets, Canonical Transformations, \CTR s, and the \PB.}
 The \PB\ \ci{poissonbrak,Becchi:1975nq,zinnbook,taylor,Zinnarticle,Weinberg2}  has the same form as a Poisson Bracket\footnote{The \PB\ is a Poisson Bracket, except that it uses Grassmann anticommuting quantities (these actually simplify things 
somewhat).} in classical mechanics  
\ci{goldstein,LandauLifmechanics}.  There is no essential distinction between coordinates and momentum for classical mechanics, but there is one for the \PB.  The reason is that the Fields are quantized and the Zinn Sources are not. Nevertheless, Canonical Transformations play a role for both kinds of Poisson Brackets, because they leave the Poisson Bracket invariant \ci{goldstein,LandauLifmechanics}.   
  For the \PB\ case we are calling these Canonical Transformations `\CTR s', because they can map one action to another, which a Canonical Transformation would never do in Classical Mechanics.
But it is important to remember that these \CTR s must always yield an  Action which yields zero for the resulting \PB, and that is because they are Canonical Transformations in their mathematical form.

\refstepcounter{orange} \label{backwards}
{\bf \theorange}.
{\bf Finding Mass and Interaction Terms for the \HCM\ by Starting with the Known Mass and Interaction Terms for the \CM:} 
So now we see that the \HCM\ arises from the \CM\ through the \CTR\ above.
This is useful because we know how to make masses and interactions for the \CM, and we did this in \pgh\ \ref{Chiraltheory2}.  Can we put those into a \CM\ and then use the \CTR\ to deduce what they look like for the \HCM\ from that?  The answer is yes!  Let us see how this works in detail,  by adding a mass term and a cubic interaction term to  (\ref{MMM}).   This is a little tricky, because the two theories are related by an \CTR\ only when they have both had their auxiliaries integrated, and the auxiliaries are different--the \CM\ has a Scalar $F$ and the \ICDSS\ has a vector auxiliary $W_{\a \dot \a}$.  When these auxiliairies are integrated then there is a \CTR\ that relates the two, and we call the \ICDSS\ with its $W$ auxiliary integrated, by the  
shorter and more descriptive name \HCM.
The  \CM\ with a mass term and an interaction term has the following Action, once the auxiliary has been integrated:

\[
{\cal A}_{\rm CM\;with \; Mass\;\&\; Interaction}
=  \int d^4 x \; \lt \{
-\f^{ \dot \a}
\pa_{\a\dot\a}
\ov\f^{\a}
+
{S} \Box {\oS} 
+
\oG\lt (
\ov\f_{\d}
C^{ \d}
+ \xp {\oS}\rt )
\rt.
\]
\[
+
\G
\lt (\f_{\dot\d}\oC^{\dot \d}
+ \xp {S}\rt )
+
  Z^{\dot \a} 
\lt (
  \pa_{\a \dot \a} S
C^{\a}
+ \xp \f_{\dot\a}\rt )
 + \ov Z^{ \a} 
\lt (
\pa_{\a \dot \a} \oS
\oC^{\dot\a}  
+ \xp
\ov \f_{\a}  
\rt ) 
\]
\be
\lt.
-
( \fr{1}{2} m_1 + g_1 S)
\f^{ \dot \a}
\f_{ \dot \a}
-
(\fr{1}{2} \om_1 + \og_1 \oS)
\ov \f^{  \a}
\ov \f_{  \a}
- 
\lt ( m_1 S + g_1 S^2   +Z_{\dot \a} \oC^{\dot \a}\rt )
\lt ( m_1 \oS +  \og_1 \oS^2  +\ov Z_{ \a} C^{ \a}  
\rt )
\rt \}
\la{MMMass}
\ee

We are using the notation in \pgh\ \ref{MajHClastnowchiral},  rather than  the notation in \pgh s \ref{Chiraltheory} and \ref{Chiraltheory2}.  This is done to agree with the  
notation in \ci{five}.

\refstepcounter{orange} \label{massandint}
{\bf \theorange}.
{\bf The \HCM\ Action with Mass and Interaction} 
It is elementary to use the \CTR\ from \pgh\ \ref{hccantran} on the expression in \pgh\ \ref{backwards}.  	The result is the Action for the massive interacting Majorana  \HCM:
\[
{\cal A}_{\rm MHC\;with \; Mass\;\&\; Interaction}
=  \int d^4 x \; \lt \{
-\f^{ \dot \a}
\pa_{\a\dot\a}
\ov\f^{\a}
+
 \fr{1}{\sqrt{2}}(J-iG)
 \Box  \fr{1}{\sqrt{2}}(J+iG)
\rt.
\]
\[
+
 \fr{1}{\sqrt{2}}(\h - i \U) 
\lt (
\ov\f_{\d}
C^{ \d}
+ \xp  \fr{1}{\sqrt{2}}(J+iG)
\rt )
+ \fr{1}{\sqrt{2}}(\h + i \U) 
\lt (\f_{\dot\d}\oC^{\dot \d}
+ \xp \fr{1}{\sqrt{2}}(J-iG)
\rt )
\]
\[
+
  Z_{\dot \a} 
\lt (
  \pa^{\a \dot \a}  \fr{1}{\sqrt{2}}(J-iG)
C_{\a}
+ \xp \f^{\dot\a}\rt )
 + \ov Z_{ \a} 
\lt (
\pa^{\a \dot \a}  \fr{1}{\sqrt{2}}(J+iG)
\oC_{\dot\a}  
+ \xp
\ov \f^{\a}  
\rt ) 
\]
\be
\lt.
-
(\fr{1}{2} m_1 + g_1 \fr{1}{\sqrt{2}}(J-iG))
\f^{ \dot \a}
\f_{ \dot \a}
-
(\fr{1}{2} \om_1 + \og_1 \fr{1}{\sqrt{2}}(J+iG))
\ov \f^{  \a}
\ov \f_{  \a}
\la{MMMass11}\ebp
- \lt ( m_1 \fr{1}{\sqrt{2}}(J-iG) +g_1 \fr{1}{2}(J-iG)^2  + Z_{\dot \a} \oC^{\dot\a}  
 \rt )
\lt ( m_1 \fr{1}{\sqrt{2}}(J+iG) +\og_1\fr{1}{2}(J+iG)^2  +\ov Z_{ \a} C^{\a}
\rt )
\rt \}
\la{MMMass2}
\ee

\refstepcounter{orange} \label{backwardsconfusion}
{\bf \theorange}.
{\bf Details for the  Majorana \HCM\ with mass, and the Loss of the SUSY Charge} 
Let us set $g_1\ra 0$   in \pgh\ \ref{massandint}.
Then  the expression in \pgh\ \ref{massandint} is the \HCM\ with just a  Majorana mass term.   The only difference from the massless case is:

\be
{\cal A}_{\rm MHC\;with \; Mass}
=  \int d^4 x \; \lt \{\cdots
\ebp
- \fr{1}{2}
m_1 
\f^{ \dot \a}
\f_{ \dot \a}
- \fr{1}{2}
m_1 
\ov \f^{  \a}
\ov \f_{  \a}
- 
\lt ( m_1  \fr{1}{\sqrt{2}}(J-iG)
 +   Z_{\dot \a} C^{\a}\rt )
\lt ( m_1  \fr{1}{\sqrt{2}}(J+iG)
 + \ov Z_{ \a} \oC^{\dot\a}  
\rt )
\rt \}
\la{MMMass3}
\ee
We see that indeed there is a mass term here for the Scalar, namely
\be
{\cal A}_{\rm MHC\;with \; Mass}
=  \int d^4 x \; \lt \{\cdots
- 
\fr{m_1^2}{2  } 
(J^2 + G^2)
\rt \}
\la{etrwggew}
\ee
But note that there is also a `mass' term for the Zinn Source $J$ in Equation (\ref{etrwggew}), 
and then there are extra terms that are all in the Zinn Action
\be
{\cal A}_{\rm MHC\;with \; Mass}
=  \int d^4 x \; \lt \{\cdots
- \lt (
   Z_{\dot \a} C^{\a}\rt )
\lt (  \ov Z_{ \a} \oC^{\dot\a}  
\rt )
\ebp- \lt (
\fr{m_1  }{\sqrt{2}}  (J-iG)
\rt )
\lt (  
 \ov Z_{ \a} \oC^{\dot\a}  
\rt )
- \lt (
   Z_{\dot \a} C^{\a}\rt )
\lt (  \fr{m_1 }{\sqrt{2}}(J+iG)
\rt )
\rt \}
\ee
So here is what we have discovered:
The \HCM\ Action has one Scalar $G$ and a Majorana Spinor $\f$ and also the Source $J$.
When we generate the \HCM\  mass term from the massive \CM\ plus the \CTR, we get a massive Spinor and a massive Scalar $G$.
We also get an object that looks like a mass term for $J$, but $J$ is a Zinn Source.

Because this is all a result of the \CTR, we are guaranteed that it will satisfy the 
\HC\ \PB\ in Equation (\ref{mhcmasterdf}).

But just looking at it we can see that it describes a Multiplet of SUSY that has one Scalar and a Spinor--but this is clearly not a proper mass multiplet that forms a representation of the SUSY algebra--that needs two Scalars and there is only one here.  And yet there is some mass degeneracy here, as though the SUSY algebra is `half-present'.

And that is the essential point of this entire paper!  The \CTR\ has enabled us to build a 
SUSY theory that  does not have a nice conserved Noether Charge--the physical theory here is not the proper one we would expect from a theory with a conserved Noether current--it has this Source $J$ where the Scalar should be.  
And because of the \CTR, it has the right set of Zinn Source terms to satisfy the \HC\ \PB.

\refstepcounter{orange} \label{withinteraction}
{\bf \theorange}.
{\bf Details for the \HCM\ with Mass and  Interactions} 
Now consider the case where $g_1 \neq 0$   in \pgh\ \ref{massandint}.
The Action there has  both mass and interactions.  Note the complicated way that the Zinn Source is intertwined with the Scalar Field.  We could do the same with a Chiral Action that also includes any other kind of interactions, with Gauge, other \CM s, even \SG.  We would end up with a lot of Zinn Source terms and an Action that only goes half-way towards a representation of the SUSY algebra.

The Action in  \pgh\ \ref{massandint} is probably the simplest possible \HC\ massive interacting theory, and it would be worth while to examine its nilpotent BRST operator $\dB$ (this is the `square root' of the \PB), and its one loop diagrams to get a feel for how this \HCM\ works at one loop.

These \HCM s will be used in \ci{five} for the Higgs Multiplets.  We will use a Dirac \HCM\ and a Majorana \HCM\ there.

  \refstepcounter{orange}\la{UCM}
  {\bf \theorange}.\;{\bf \UCM s}
If we take the \CTR\ that goes all the way, to generate an \UCM, we get a theory with no Scalar and some interesting Zinn terms, and a mass just for the Spinor. That case is actually simpler than the above, and we write it down in \pgh\ \ref{massforUCM}.
In that case the interaction term would just add to the Zinn Source sector, and the theory would  still be a free massive theory.  To get interactions there requires Gauge theory.

The \CTR\ can be applied twice, so that both Scalars are removed from the Lagrangian.  The new Lagrangian gains two terms with Antighosts while the Fermions remain as quantized Fields. 
 Start with the \CM\ with the Action in Equation (\ref{MMM}).
Now  consider
using an \CTR\ generated as follows.
Instead of the \CTR\ in \pgh\ \ref{hccantran} we now note that  the old `Fields' 
 $(S,\oS)$ are conjugate to the  old
`Zinn Sources' $(\G,\ov \G)$, and we want an \CTR\ that takes us to the new  `Fields' $(\h,\ov \h)$ which are conjugate to the 
new `Zinn Sources'  $(J, \oJ)$.
We choose a generating  functional  of the new Zinn Sources  $(J,\oJ)$ and the old `Zinn Sources' $(\G,\ov \G)$.
This is
\be
\cG_{\rm  MUC} = 
\int d^4 x \lt \{
\G J
+
\oG \oJ 
 \rt \}
\ee
and we have
\be
S \ra \fr{\d \cG}{\d \G}
=J;\;\oS \ra  \fr{\d \cG}{\d \ov\G}=
\oJ;\;
\ee
and
\be
\G  = \fr{\d \cG}{\d J}
\ra \h;\;\ov \G =  \fr{\d \cG}{\d \oJ}\ra
\ov \h;\;
\ee

We get the following transformed Action:
\[
{\cal A}_{\rm MUC}
=  \int d^4 x \; \lt \{
-\f^{ \dot \a}
\pa_{\a\dot\a}
\ov\f^{\a}
+
{J} \Box {\oJ} 
+
\ov \h \lt (
\ov\f_{\d}
C^{ \d}
+ \xp \oJ\rt )
+
\h
\lt (\f_{\dot\d}\oC^{\dot \d}
+ \xp {J}\rt )
\rt.
\]
\be
\lt.
+
  Z_{\dot \a} 
\lt (
  \pa^{\a \dot \a} J
C_{\a}
+ \xp \f^{\dot\a}\rt )
 + \ov Z_{ \a} 
\lt (
\pa^{\a \dot \a} \oJ
\oC_{\dot\a}  
+ \xp
\ov \f^{\a}  
\rt ) 
- 
   Z_{\dot \a} C^{\a}
\ov Z_{ \a} \oC^{\dot\a}  
\rt \}
\la{newuncactmaj}\ee

which yields zero for the new \PB
\be
\cP_{\rm MUC}[\cA]
=
\int d^4 x \lt \{
\fr{\delta \cA}{ \delta Z^{\dot \a}}
\fr{\delta \cA}{ \delta \f_{\dot \a}}
+
\fr{\delta \cA}{ \delta \oZ^{  \a}}
\fr{\delta \cA}{ \delta \ov \f_{ \a}}
+
\fr{\delta \cA}{ \delta  \ov\h}
\fr{\delta \cA}{ \delta \oJ }
+\fr{\delta \cA}{ \delta  J}
\fr{\delta \cA}{ \delta \eta }
\rt \}+\fr{\pa \cA}{ \pa h_{\a\dot \a} }
\fr{\pa \cA}{ \pa \x^{\a\dot \a} }
\ee

This looks very similar to  (\ref{MMM}), but the theory is not at all the same. The $J$ are not quantized and the $\h$ are quantized. So the complex quantized Scalar Field $(S,\oS)$ is gone from the theory along with its Zinn Source $(\G,\ov \G)$, while the quantized Fermion $(\f, \ov \f)$ remains, and  the new Zinn Sources $(J,\oJ)$ and quantized Antighosts $(\h,\ov \h)$ appear.  We are assured that the new Action satisfies the new \PB, because the old Action satisfied the old \PB.

This procedure does not correspond to any known starting Action like the \HCM\ discussed above in section \ref{MHCM}. Equation (\ref{newuncactmaj}) is a new Action.
We started with the \HCM s found by BRST recycling, and then found the \CTR s that took those theories to \CM s. Now by  generalizing those \CTR s we have discovered new theories that have no physical Scalars at all, just physical Fermions.

\refstepcounter{orange}\la{massforUCM}
{\bf \theorange}.\;{\bf  Mass Term for the \UCM}
In \pgh\ \ref{backwardsconfusion} above, we discussed the mass term for the \HCM.  It has a mass term that only goes half-way towards that of a \CM. 
Now let us  start again with the \CM\ with the Action in Equation (\ref{MMM}), with a mass term, so that we get Equation (\ref{MMMass}), and then use 
the \CTR\ in \pgh\ \ref{UCM} on  Equation (\ref{MMMass}). This yields the following Action (we are setting $g_1\ra 0$):
\[
{\cal A}_{\rm Majorana\; UnChiral\;Massive}
=  \int d^4 x \; \lt \{
-\f^{ \dot \a}
\pa_{\a\dot\a}
\ov\f^{\a}
+
{J} \Box {\oJ} 
+
\ov \h \lt (
\ov\f_{\d}
C^{ \d}
+ \xp \oJ\rt )
+
\h
\lt (\f_{\dot\d}\oC^{\dot \d}
+ \xp {J}\rt )
\rt.
\]
\[
+
  Z_{\dot \a} 
\lt (
  \pa^{\a \dot \a} J
C_{\a}
+ \xp \f^{\dot\a}\rt )
 + \ov Z_{ \a} 
\lt (
\pa^{\a \dot \a} \oJ
\oC_{\dot\a}  
+ \xp
\ov \f^{\a}  
\rt ) 
\]
\be
\lt.
-
m_1 \fr{1}{2}
\f^{ \dot \a}
\f_{ \dot \a}
-
m_1 \fr{1}{2}
\ov \f^{  \a}
\ov \f_{  \a}
- 
\lt ( m_1 J +   Z_{\dot \a} C^{\a}\rt )
\lt ( m_1 \oJ + \ov Z_{ \a} \oC^{\dot\a}  
\rt )
\rt \}
\la{MMMassusc}
\ee
The Scalar Mass terms ${S} \lt ( \Box  -m_1\om_1 \rt )  \oS$ in Equation (\ref{MMMass}) 
 have become the  Zinn Source term $J \lt ( \Box  -m_1\om_1 \rt )  \oJ$ in Equation 
(\ref{MMMassusc}).
So  Equation 
(\ref{MMMassusc})  is an Action which has a Fermionic Mass term and no Scalars.  
Note that it is quite a lot simpler than the \HCM.
We are guaranteed that it will satisfy the appropriate \PB\ for the \UCM, and so SUSY is still preserved.  But the SUSY Charge has been suppressed in this 
\UCM\ Quantum Field theory.

It is clear that this is not a representation of the SUSY algebra on the physical states.

This is the kind of SUSY multiplet that we will use for the Quarks and Leptons in
\ci{five}, except that we need to have the Dirac version for that.  A similar exercise to the above would show that there is only a massive Fermion left in the Dirac Unchiral Multiplet, if one adds a Dirac type mass term to  \pgh\ \ref{diracpairChiralmults}.

\refstepcounter{orange}\la{derivenewbrstforCM}
{\bf \theorange}.\;{\bf 
A Derivation of the \PB\ for the Interacting \CM:}
 Let us recall the derivation of the \PB\ for the \CM\ interacting with \SYM.  
Some examples of these are in \ci{five} and the notation here is adapted to the (1,0) Multiplet used there, but we do not need much detail for this.  Let us start with a \CM\ Action like that in Equation \ref{MMMass}, except with Gauge interactions and perhaps more \CM s too.
To derive the \PB\ identity, one starts with an Action
\be
\cA_{\rm CM}[\f,\ov \f,S,\oS,Z,\ov Z, \G, \oG;{\rm Gauge\;Fields\;\&\;
Gauge\; Zinn\; Sources}]
\la{orignte}\ee
and writes the Feynman path integral over all the fields, with Sources $j_S,  \ov j_S, j_{\f}, \ov j_{\f}$ for the Fields and `Zinn Sources' for the variations of the Fields:
\[
\cG_{\rm CM}[Z,\ov Z, \G, \oG, j_S,  \ov j_S, j_{\f}, \ov j_{\f};{\rm Gauge\;Terms}]
\]
\be
= 
   \int \d \f \d \ov \f \d S \d \ov S;\d  {\rm Gauge}\;
e^{i \lt \{ \cA_{\rm CM} + \int d^4 x \lt ( j_S S + j_{\f} \f + *+ {\rm Gauge\;Terms} \rt ) \rt \} }
\ee
So we have Scalar Fields $S$ and Fermion Fields $\f$ here, and also 
Gauge field integration variables with a form like:
\be
\d \lt( {\rm Gauge}\rt )= \d \c \d \ov\c \d V \d D \d \w \d \z
\ee
where the variables are Gauge and Ghost Fields, as defined in \ci{one}.
Then one shifts all the fields as follows ($\lam$ is an anticommuting parameter so that the Grassmann nature of the fields is not violated.)
\be
S \ra S + \lam
\fr{\d  \cA_{\rm CM}}{\d \G}
;\; \f  \ra \f + \lam
\fr{\d  \cA_{\rm CM}}{\d Z}
\ee
plus \CC s, and 
with similar terms for the Gauge Fields.
We assume that the $F$ type auxiliaries have been integrated, as in Theorem 1 above.
Now since, for the Chiral Theory, we have
\be
\d_{\rm CM\; Fields} \cA_{\rm CM}=0
\ee
where
\be
\d_{\rm CM\;Fields} 
=
\int d^4 x \lt \{
\fr{\delta \cA_{\rm CM}}{ \delta Z^{\dot \a}}
\fr{\delta  }{ \delta \f_{\dot \a}}
+
\fr{\delta \cA_{\rm uCM}}{ \delta \oZ^{  \a}}
\fr{\delta  }{ \delta \ov \f_{ \a}}
+
\fr{\delta \cA_{\rm uCM}}{ \delta \G }
\fr{\delta  }{ \delta  S}
+
\fr{\delta \cA_{\rm uCM}}{ \delta \ov \G }
\fr{\delta  }{ \delta  \ov S}
\rt \} 
+ \d_{\rm Gauge\;Fields} + C \oC \fr{\pa}{\pa \x}
\ee
this yields 
\[
\cG_{\rm CM}[Z,\ov Z, \G, \oG, j_S,  \ov j_S, j_{\f}, \ov j_{\f}; {\rm Gauge\;Terms}]]= \]
\[
 \int \d \f \d \ov \f \d S \d \ov S \;\d {\rm Gauge}\;
e^{i \lt \{ \cA_{\rm CM} + \int d^4 x \lt ( j_S S + j_{\f} \f + *\rt ) \rt \} } 
\int d^4 y \lt \{( j_S \d S - j_{\f} \d \f + * + {\rm Gauge\;Terms}\rt \}
\]
\be
 \equiv
\lt <\int d^4 y \lt \{( j_S \d S - j_{\f} \d \f + * + {\rm Gauge\;Terms}\rt \} \rt > =0
\ee
We follow the derivation of Zinn-Justin \ci{zinnbook} by defining the Generator of Connected Diagrams:
\be
\cG_{\rm CM}=  e^{i \cG_{\rm Connected}}
\ee
and the Generator of One-Particle-Irreducible Diagrams:
\be
\cG_{\rm Connected} = \cG_{\rm 1PI} + 
\int d^4 y \lt \{( j_S  S - j_{\f}  \f + *+ {\rm Gauge\;Terms} \rt \}
\ee
where the Legendre transform is defined by:
\be
\fr{\d \cG_{\rm 1PI}}{\d S} =
 j_S ;\;
\fr{\d \cG_{\rm 1PI}}{\d \f} =
 -j_{\f} 
\ee
and the Zinn Sources bring in variations
\be
\fr{\d \cG_{\rm Conn}}{\d \G} =
\fr{\d \cG_{\rm 1PI}}{\d \G} \equiv
 \lt < \d S \rt >
\ee

\be
\fr{\d \cG_{\rm Conn}}{\d Z} =
\fr{\d \cG_{\rm 1PI}}{\d Z} \equiv
 \lt < \d \f \rt >
\ee

and their \CC s plus  {\rm Gauge\;Terms}.  So we get\footnote{\la{supprob} There is a problem with the $\x$ terms here, that can be solved by coupling to \SG\ as explained in footnote \ref{coupletoSG} on page \pageref{coupletoSG}. }  the \PB.
\[
\cP_{\rm CM}[\cA]
=
\int d^4 x \lt \{
\fr{\delta \cG_{\rm 1PI}}{ \delta Z^{\dot \a}}
\fr{\delta \cG_{\rm 1PI}}{ \delta \f_{\dot \a}}
+
\fr{\delta \cG_{\rm 1PI}}{ \delta \oZ^{  \a}}
\fr{\delta \cG_{\rm 1PI}}{ \delta \ov \f_{ \a}}
+
\fr{\delta \cG_{\rm 1PI}}{ \delta  \G}
\fr{\delta \cG_{\rm 1PI}}{ \delta S }
\rt.\]
\be
\lt.+\fr{\delta \cG_{\rm 1PI}}{ \delta  \ov\G}
\fr{\delta \cG_{\rm 1PI}}{ \delta \ov S}
\rt \}+{\rm Gauge\;Part}+ \fr{\pa \cG_{\rm 1PI}}{ \pa h_{\a\dot \a} }
\fr{\pa \cG_{\rm 1PI}}{ \pa \x^{\a\dot \a} }
=0\ee
which summarizes all the BRST identities here.

\refstepcounter{orange}\la{derivenewbrstforUCM}
{\bf \theorange}.\;{\bf 
A Derivation of the \PB\ for the Interacting \UCM\ that arises from the Above \CM:}
  Here we  derive the  identity for the \UCM.   Exactly the same kind of reasoning applies to \HCM s, but the notation is more complicated for that case.
This is similar to the derivation in \pgh\ \ref{derivenewbrstforCM}, but it has one step that is not immediately obvious.
  The \UCM\ is defined by simply changing variables in the \CM\ in \pgh\ \ref{derivenewbrstforCM}:
\[
\cA_{UCM}
[\f,\ov \f,J,\ov J,Z,\ov Z, \h, \ov \h;{\rm Gauge\;Fields\;\&\;
Gauge\; Zinn \;Sources}]
\]
\be
=
\cA_{CM}[\f,\ov \f,S\ra J,\ov S\ra \ov J,Z,\ov Z, \G\ra \h, \oG\ra \ov \h;{\rm Gauge\;Fields\;\&\;
Gauge\; Zinn\; Sources}]
\la{orignteder}
\ee
Now we want to derive a new \PB\ using this Action.  We have a different set of variables in the  Feynman path integral over all the fields, because $S$ is gone and $\h$ has appeared:
\[
\cG_{UCM}[Z,\ov Z, j_{\h}, \ov j_{\h} , J, \oJ,j_{\f},  \ov j_{\f};{\rm Gauged}]=
   \int \d \f \d \ov \f \d \h \d \ov \h; \d \lt( {\rm Gauge}\rt ) 
\]
\be
e^{i \lt \{ \cA_{\rm UCM} + \int d^4 x \lt ( j_{\h} \h + j_{\f} \f + * + {\rm Gauge\;Terms}\rt ) \rt \} }
\ee

So now the integration variables are the Scalar Antighost Fields $\h$ and Fermion Fields $\f$, and also 
Gauge field integration variables.
Now we do the usual shift of fields
\be
\h \ra \h + \lam
\fr{\d  \cA_{\rm UCM}}{\d J}
\ee

\be
\f  \ra \f + \lam
\fr{\d  \cA_{\rm UCM}}{\d Z}
\ee
plus \CC s
with similar terms for the Gauge Fields.
Now  for the moment for the \UCM, we  assume the following:
\be
\d_{\rm Fields} \cA_{\rm UCM}=0
\la{invofucm}
\ee
{\bf This is the crucial point, and we will return to this below in \pgh\ \ref{derivefieldtransforUCM}}.
Now
we simply shift all the fields in the integrand, with the result
\[
    \int \d \f \d \ov \f \d \h \d \ov \h; \d \lt( {\rm Gauge\;Fields}\rt )
e^{i \lt \{ \cA_{\rm UCM} + \int d^4 x \lt ( j_{\h} \h + j_{\f} \f + *+ {\rm Gauge\;Terms}\rt ) \rt \} }
\]
\be
\int d^4 y \lt \{( j_{\h} \d \h + j_{\f} \d \f + *+ {\rm Gauge\;Terms} \rt \}=0
\ee

and defining as usual
\be
\cG_{\rm UCM}= 
e^{i \cG_{\rm Connected}}
\ee
and 
\be
\cG_{\rm Connected} = \cG_{\rm 1PI} + 
\int d^4 y \lt \{( j_{\h}  \h + j_{\f}  \f + * +  {\rm Gauge\;Terms}\rt \}
\ee
where
\be
\fr{\d \cG_{\rm 1PI}}{\d \h} =
- j_{\h} 
;\;\fr{\d \cG_{\rm 1PI}}{\d \f} =
- j_{\f} 
\ee
and their \CC s, plus  {\rm Gauge\;Terms}, we
 get\footnote{The same problem with the $\x$ terms that was mentioned in footnote \ref{supprob} on page \pageref{supprob} occurs here  again, and again it can be solved by coupling \SG\ to these new Multiplets as explained in footnote \ref{coupletoSG} on page \pageref{coupletoSG}  }  the \PB\ just as we did above in \pgh\ \ref{derivenewbrstforCM}:
\[
\cP_{\rm UCM}[\cG_{\rm 1PI}]
=
\int d^4 x \lt \{
\fr{\delta \cG_{\rm 1PI}}{ \delta Z^{\dot \a}}
\fr{\delta \cG_{\rm 1PI}}{ \delta \f_{\dot \a}}
+
\fr{\delta \cG_{\rm 1PI}}{ \delta \oZ^{  \a}}
\fr{\delta \cG_{\rm 1PI}}{ \delta \ov \f_{ \a}}
\rt.
\]
\be
\lt.
+
\fr{\delta \cG_{\rm 1PI}}{ \delta  \h}
\fr{\delta \cG_{\rm 1PI}}{ \delta J }
+\fr{\delta \cG_{\rm 1PI}}{ \delta  \ov\h}
\fr{\delta \cG_{\rm 1PI}}{ \delta \ov J}
\rt \}+{\rm Gauge\;Part}+ \fr{\pa \cG_{\rm 1PI}}{ \pa h_{\a\dot \a} }
\fr{\pa \cG_{\rm 1PI}}{ \pa \x^{\a\dot \a} }=0
\ee
This shows that this Action also generates a supersymmetric theory, although there are plenty of questions that we need to understand about the nature of that theory.

\refstepcounter{orange}\la{derivefieldtransforUCM}
{\bf \theorange}.\;{\bf The Invariance Equation $\d_{\rm Fields} \cA_{\rm UCM}=0
$ :}
The only step that was tricky in the above was Equation (\ref{invofucm})
\be
\d_{\rm Fields} \cA_{\rm UCM}=0
\ee
This is important, because if we needed to also vary the Zinn Sources in the Action  $\cA_{\rm UCM}$, we could not derive the \PB, because of course Zinn Sources are not integrated in the \Fpi.  How can we know this with so little detail being shown here?

We know from \pgh\ \ref{derivenewbrstforCM} that;
\[
\cP_{\rm CM}[\cA_{\rm CM}]
=
\int d^4 x \lt \{
\fr{\delta \cA_{\rm CM}}{ \delta Z^{\dot \a}}
\fr{\delta \cA_{\rm CM}}{ \delta \f_{\dot \a}}
+
\fr{\delta \cA_{\rm CM}}{ \delta \oZ^{  \a}}
\fr{\delta \cA_{\rm CM}}{ \delta \ov \f_{ \a}}
\rt.
 \]
\be
\lt.
+
\fr{\delta \cA_{\rm CM}}{ \delta  \G}
\fr{\delta \cA_{\rm CM}}{ \delta S}
+\fr{\delta \cA_{\rm CM}}{ \delta  \ov\G}
\fr{\delta \cA_{\rm CM}}{ \delta \ov S}
\rt \}+{\rm Gauge\;Part}+ \fr{\pa \cA_{\rm CM}}{ \pa h_{\a\dot \a} }
\fr{\pa \cA_{\rm CM}}{ \pa \x^{\a\dot \a} }=0
\ee
and hence, because they are related by the \CTR: 
\[
\cP_{\rm UCM}[\cA_{\rm UCM}]
=
\int d^4 x \lt \{
\fr{\delta \cA_{\rm UCM}}{ \delta Z^{\dot \a}}
\fr{\delta \cA_{\rm UCM}}{ \delta \f_{\dot \a}}
+
\fr{\delta \cA_{\rm UCM}}{ \delta \oZ^{  \a}}
\fr{\delta \cA_{\rm UCM}}{ \delta \ov \f_{ \a}}
+
\fr{\delta \cA_{\rm UCM}}{ \delta  \h}
\fr{\delta \cA_{\rm UCM}}{ \delta J }
\rt.
\]
\be
\lt.+\fr{\delta \cA_{\rm UCM}}{ \delta  \ov\h}
\fr{\delta \cA_{\rm UCM}}{ \delta \ov J}
\rt \}+{\rm Gauge\;Part}+ \fr{\pa \cA_{\rm UCM}}{ \pa h_{\a\dot \a} }
\fr{\pa \cA_{\rm UCM}}{ \pa \x^{\a\dot \a} }=0
\ee
But the latter can be written in the form:
\be
\d_{\rm UCM\;Fields} \cA_{\rm UCM}=0
\la{fsdefsf}\ee
where
\[
\d_{\rm UCM\;Fields} 
=
\int d^4 x \lt \{
\fr{\delta \cA_{\rm UCM}}{ \delta Z^{\dot \a}}
\fr{\delta  }{ \delta \f_{\dot \a}}
+
\fr{\delta \cA_{\rm UCM}}{ \delta \oZ^{  \a}}
\fr{\delta  }{ \delta \ov \f_{ \a}}
+
\fr{\delta \cA_{\rm UCM}}{ \delta J }
\fr{\delta  }{ \delta  \h}
+
\fr{\delta \cA_{\rm UCM}}{ \delta \oJ }
\fr{\delta  }{ \delta  \ov \h}
\rt \} 
\]
\be
+ \d_{\rm Gauge\;Fields} + C \oC \fr{\pa}{\pa \x}
\ee
The point is that one can view the equation (\ref{fsdefsf}) either as a variation of the fields or the Zinn Sources or any combination, but there is no need to include both.    So the fact that the original Action in Equation (\ref{orignte}) is invariant under transformations of the fields alone, means that the derived Action 
in Equation (\ref{orignteder})
 is invariant under transformations of the fields alone, even though they are different fields in the two cases. 
 Exactly the same kind of reasoning applies to \HCM s.

\refstepcounter{orange}\la{hamsection}
{\bf \theorange}.\;{\bf Remarks about the Hamiltonian and the Cohomology:}
In \ci{Witten:1981nf}, for example, it was shown that the Hamiltonian is the sum of the SUSY charges $H = \sum Q_i^2$ and that the conservation of the SUSY charge implies that 
$\dot Q_i = \lt [ H, Q_i\rt ]=0$.  These imply that there is a set of states with the same mass and different spins. So how can these `Suppressed SUSY Charge' theories exist without this structure?

The present mechanism gets around this reasoning because it is simply impossible to construct a SUSY charge $Q_i$ in these new theories.  Nevertheless, a large amount of the power of SUSY is maintained by the fact that the new \PB\ is true for the new theories.  Furthermore, the algebra of the Chiral SUSY transformations (acting on the Zinn Sources as well as the Fields, in the new theories) is still used to satisfy the new \PB\ at the Lagrangian level.  

It appears to be,  more or less, a coincidence that  the Chiral Action and the  \HC\ Action are related by cohomology considerations\footnote{This is rather like the derivation of the new transformations in \ci{one} from the notion of `BRST Recycling'--there are a limited number of ways to construct these actions, and so things `pop up again' in unexpected places.}.  These cohomology considerations come from  the nilpotent BRST cohomology operator $\dB$ which is the `square root' of the \PB.  There might be a deeper connection and a clearer explanation, but, if those explanations exist, they are presently quite obscure to the author.

  It is much easier to simply \ul{\bf use} the mechanism of `Suppressed SUSY Charge', by noting that the SUSY charge is not present in some parts of the new actions, and that the new actions are constrained by the new \PB, than it is to explain `\ul{\bf why}' this mechanism exists. It just does exist, and it is simple to use,  and its use solves many outstanding problems of the SSM, as will be seen in \ci{five}.

\refstepcounter{orange}\la{conclusion2}
{\bf \theorange}.\;{\bf  Conclusion:}
  In this paper we have shown that, for any   
chosen \CM, one can easily derive and write down three more theories\footnote{For a Dirac Multiplet this is limited by the conserved global U(1) phase, which must be conserved.}.  The chosen \CM\  can be coupled to SUSY Gauge theory and other \CM s, and the interactions of the new theories follow directly and simply from those interactions.   But the three new Actions are quite different in their behaviour from the original chosen one.    This can be done for each \CM\ in the theory  
independently.

It is in the Chiral form that it is easy to write down the couplings.  Then one  simply implements the appropriate set of \CTR s, which result in a new Action and a new \PB.  The new Action yields zero for the new \PB.  This {Exchange} Transformation takes all or part of the Scalar Fields $S$ and replaces them with Zinn Sources $J$, and also takes the related Zinn Sources $\G$ and replaces them with Antighost Fields $\h$.
These \CTR s are closely related to Canonical Transformations, as is explained in \pgh\
\ref{aboutCTRs}.

These new Actions were discovered by generating the \ICDSS\ in \ci{one}, in the hope of coupling the BRST cohomology of the \CM\ to it.  Once the auxiliary $W$ was integrated in that theory, it was noticed that the resulting \HC\ Action could be generated by an \CTR\ from a Chiral Action that has its auxiliary F integrated.  So, in a sense, that coupling of the cohomology has now been done, and the result is that we have discovered new ways to realize SUSY in local Actions, and those new \HC\ Actions can be coupled to Gauge theory and each other (and even \SG). 

But the result here goes farther, because we also have discovered \UCM s that do not arise by way of the \ICDSS\ of \ci{one}.  The \UCM\ Action arises simply by taking the full version of the \CTR\ that was suggested by the existence of the \HCM.

Supercharges are not constructible in the new theories (except in some sectors), because
when the Zinn Sources become involved, no divergenceless SUSY current exists.  This happens because the Zinn sources are not quantized, and they do not satisfy Equations of Motion.

The new theory has new Fields and Sources, and so it has a different generating functional for the \Fpi, as shown in \pgh s \ref{derivenewbrstforCM} and \ref{derivenewbrstforUCM}.  But because the Actions and \PB s are  related through an \CTR, it was easy to prove, in \pgh s \ref{derivenewbrstforUCM} and \ref{derivefieldtransforUCM}  that the new Action satisfies the new 
\PB\ for its 1PI Generating Functionals $\cG_{\rm 1PI}$.

In  \pgh s
\ref{backwards},  \ref{massandint} and 
\ref{backwardsconfusion} we set out the details for the massive \HCM\ case, which shows explicitly how the effect of a SUSY Charge acts like it is `half-present'.
In \pgh\ \ref{withinteraction} we observed the interactions of the \HCM.

 \pgh\ \ref{massforUCM} discusses mass for the \UCM\ case and shows that the SUSY Charge is completely gone there.  In that case we need to couple the theory to SUSY Gauge Theory to get an interaction.

The result is that that these new theories, when calculated in the renormalized Feynman expansion, iteratively, loop by loop, should be as valid  as is the original purely Chiral theory  \ci{dixonnucphys}.   But the theories are very different, of course, and the \HC\ and \UC\ theories are not subject to the SUSY algebra, as described in \ci{algebrarefs}, because they do not have conserved SUSY Charges.  They are lacking in Scalar Fields compared to \CM s. We shall use some of these new \HCM s and \UCM s, coupled to SUSY Gauge theory and to each other,  to write down a new kind of SSM,  in \ci{five}.

In the Appendix at \pgh\ \ref{diraccase}, we review the Dirac case.  This is necessary to describe  Leptons and Quarks, since they have conserved Baryon and Lepton numbers.  
This is not very different from the Majorana case which is discussed above.

\begin{center}
 {\bf Acknowledgments}
\end{center}
\vspace{.2in}

  I thank  Carlo Becchi, Friedemann Brandt, Cliff Burgess, Philip Candelas, Rhys  Davies, James Dodd, Mike Duff,  Chris Hull, Pierre Ramond, Peter Scharbach,   Kelly Stelle, 
J.C. Taylor and Peter West for stimulating correspondence and conversations.  Raymond Stora introduced me to spectral sequences, which eventually brought  \cdss s forward, and he will be missed very much. I particularly want to thank 
Xerxes Tata who took the trouble to insist on the importance of the tachyon problem for the \cdss, and also on the questions surrounding the non-conservation of the Supercharge. These were the origin of the \HCM, which gave rise in turn to the \CTR, which is the foundation of this paper and its companion on the SSM. Some of this work was done at the Mathematical Institute at Oxford University, and  I thank Philip Candelas and the Mathematical Institute for hospitality.

\begin{center}
 {\bf Appendix: The  Dirac \ICDSS, \\ and  the Dirac \HCM}
\end{center}
\vspace{.2in}

\refstepcounter{orange} \label{diraccase}
{\bf \theorange}.
{\bf The  Dirac \ICDSS}: 
The Majorana \ICDSM\ treated above is the Majorana version of the work
in \ci{one,two,three}.  The progress reported there was to find and use  the
following free massless kinetic Action for a generic Dirac (ie Complex)
\ICDSS\footnote{Go to reference \ci{one}, and change $E\ra
\sqrt{2} G, \ov \U \ra \fr{1}{\sqrt{2}} \ov \U$
and $\h\ra \sqrt{2} \h, J \ra \fr{1}{\sqrt{2}} J$ and change
overall sign of kinetic term.}:
\be
{\cal A}_{\rm DI}
={\cal A}_{\rm DI\;Kinetic}
+
\cA_{\rm DI\; Zinn}
+
\cA_{\rm SUSY}
\ee

      \refstepcounter{orange} {\bf \theorange}\;
      The  Dirac \ICDSS\ has the following kinetic Action.  We will use the notation of  \ci{one}.  Here we also will add a group
representation\footnote{Here is the rule governing the index:
\emph{Index Rule: All the left objects have index up, and all
the right objects have the index down. The reverse rule holds
for the \CC, and the \CC\ of left is right and vice versa. In
special cases there may also be a way to raise and lower the
index with some invariant tensor.}} index $i$:
   \[
{\cal A}_{\rm DI\;Kinetic}
= \int d^4 x \; \lt \{
-\f_{L}^{i\dot \a}
\pa_{\a\dot\a}
\ov\f_{Li}^{ \a}
-
\f_{Ri}^{\dot \a}
\pa_{\a\dot\a}
\ov\f_{R}^{i \a}
-
 W^i_{\a\dot \a}
\ovW_i^{\a\dot \a}
+
H^i\Box \ov H_i 
\rt.\]
\be-
 \fr{ \sqrt{2} }{2}  
\ov \eta_{i} \lt (
 \f_L^{i\dot \d}
\oC_{\dot \d}
+
\ov\f_{R}^{i\d}
C_{ \d}
\rt )
\lt.
-\fr{ \sqrt{2} }{2}
 \eta^{i}  \lt (
 \ov\f_{Li}^{  \d}
C_{\d}
+
 \f_{Ri}^{\dot\d}
\oC_{\dot \d}
\rt )
\rt \}
\la{actphi}\ee

\refstepcounter{orange}{\bf \theorange}.\;
Here is the Zinn Action that goes with this Action, also taken
from \ci{one}, with indices added.
\[
{\cal A}_{\rm DI\; Zinn}
\la{zinnphi}=\int d^4 x
 Z_{Ri}^{\dot \a}
\lt (
- \fr{1}{ \sqrt{2} } \pa_{\a \dot \a} H^i
C^{\a} 
-
W^i_{\a \dot \a} C^{\a}
+
\x \cdot \pa\;\f_{L  \dot \a} 
\rt ) 
\]
\[
+
 Z_{L}^{i\dot \a}
\lt (
\fr{1}{ \sqrt{2} } \pa_{\a \dot \a} \oH_{i}
C^{\a} 
-
\ovW_{i\a \dot \a} C^{\a}
+
\x \cdot \pa\;\f_{R i \dot \a} 
\rt ) 
+
\S^{i\a\dot \a}
\lt (
 \sqrt{2}  \ov \eta_i   \oC_{\dot \a}
C_{\a}
-\fr{1}{2}
\pa_{\a}^{\dot \g} \f_{R i \dot \g} 
 \oC_{\dot \a}
\rt.
\]
\[
\lt.
-\fr{1}{2}
\pa_{\a}^{\dot \g}  \f_{R i \dot \a}
 \oC_{\dot \g}
-\fr{1}{2}\pa_{\dot\a}^{  \g} 
\ov \f_{L  i  \g} 
 C_{  \a}
-\fr{1}{2}
\pa_{\dot\a}^{ \g}  
\ov \f_{L i \a}
 C_{\g}
+ \x \cdot \pa\;
 \ovW_{i\a\dot \a}\rt )
\]
\be 
+
 \ov \U_i
\lt ( 
\fr{1}{ \sqrt{2} } \ov\f^i_{R \b}
C^{ \b}
-  
\fr{1}{ \sqrt{2} }
 \f^i_{L \dot \b}
\oC^{\dot \b}
+\x \cdot \pa G^i
\rt ) 
+
\fr{1}{ \sqrt{2} }
J^i
\lt (
\pa_{\g  \dot \d} 
 \ovW_i^{\g  \dot \d}  
  + \x \cdot \pa \ov \eta_i \rt )
+*
+\cA_{\rm SUSY}
\ee

\refstepcounter{orange}{\bf \theorange}.\;
Here is the \PB\ that goes with this Action, also taken from
\ci{one}, with indices added.
\[
\cP_{\rm DI}[\cA]=
\int d^4 x \lt \{
\fr{\d \cA}{ \d Z_L^{i\dot \a}}
\fr{\d \cA}{ \d \f_{Ri\dot \a}}
+
\fr{\d \cA}{ \d Z_{Ri}^{\dot \a}}
\fr{\d \cA}{ \d \f^i_{L\dot \a}}
+
\fr{\d \cA}{ \d \ov\S_i^{\a\dot \a}}
\fr{\d \cA}{ \d W^i_{\a\dot \a}}
\rt.
\]
\be
\lt.+
\fr{\d \cA}{ \d \ov \U_i}
\fr{\d \cA}{ \d H^i}
+\fr{\d \cA}{ \d \ov J_i}
\fr{\d \cA}{ \d \eta^i}
\rt \} +*
+\fr{\pa \cA}{ \pa h_{\a\dot \a} }
\fr{\pa \cA}{ \pa \x^{\a\dot \a} }
\la{phipb}
\ee

\refstepcounter{orange} \la{CHCMAction2}
{\bf \theorange}.\;
{\bf Integrate the Auxiliary $W$ to form the \HCM\ Action}: 
There is little difference between this Dirac version and the Majorana version analyzed above, except that the Dirac version can admit a conserved quantum number like Lepton number.  
After integrating the auxiliary out, the resulting Action can be written in the  form:

\[
\cA_{\rm DHC}
= \int d^4 x \; \lt \{
-\f_{L}^{\dot \a}
\pa_{ \a\dot\a}
\ov\f_{L}^{ \a}
-
\f_{R}^{\dot \a}
\pa_{ \a\dot\a}
\ov\f_{R}^{ \a}
+
\fr{1}{ 2 }\pa_{\a \dot \a}H \pa^{\a \dot \a}
 \ov H 
+\fr{1}{ 2 }
\pa_{\a  \dot \a} J
\pa^{\a  \dot \a}
\oJ
\rt.
\]
\[
+
\ov \eta \lt (
 \fr{1}{ \sqrt{2} } \f_{L \dot \d}
\oC^{\dot \d}
+\fr{1}{ \sqrt{2} }  
\ov\f_{R \d}
C^{ \d}
\rt )
+ \eta  \lt (
 \fr{1}{ \sqrt{2} } \ov\f_{L   \d}
C^{\d}
+\fr{1}{ \sqrt{2} }  
 \f_{R \dot\d}
\oC^{\dot \d}
\rt )
\]
\[
+
 \ov \U
\lt ( 
\fr{1}{ \sqrt{2} }  \ov\f_{R \b}
C^{ \b}
- \fr{1}{ \sqrt{2} }
 \f_{L \dot \b}
\oC^{\dot \b}
\rt )
+
 \U
\lt ( 
\fr{1}{ \sqrt{2} }  \f_{R\dot \b}
\oC^{\dot \b}
- \fr{1}{ \sqrt{2} }
 \ov \f_{L  \b}
C^{\b}
\rt )
\]

\[
+
 \oZ_{R \a}
 \oC_{\dot \a}
\lt (-\fr{1}{ \sqrt{2} }
 \pa_{ \a \dot \a} \ov H
+
\fr{1}{ \sqrt{2} }
\pa^{\a  \dot \a}
\oJ
\rt)
+
 \oZ_{L}^{  \a}
  \oC^{\dot\a}
 \lt (\fr{1}{ \sqrt{2} }
\pa_{\a \dot \a} H
+\fr{1}{ \sqrt{2} }
\pa^{\a  \dot \a}
J
\rt ) 
\]
\[
+ Z_{R}^{\dot \a}
C^{\a} 
\lt (
-\fr{1}{ \sqrt{2} }
 \pa_{ \a \dot \a} H
+\fr{1}{ \sqrt{2} }
\pa_{\a  \dot \a} J
\rt ) 
+
 Z_{L}^{\dot \a}
C^{\a}
 \lt (\fr{1}{ \sqrt{2} }
\pa_{\a \dot \a} \ov H
+\fr{1}{ \sqrt{2} }
\pa^{\a  \dot \a}
\oJ
\rt ) 
\]
\be
+
 Z_{L \dot \a}
  C_{\a}
 \oZ_{L}^{  \a}
  \oC^{\dot\a}
 +
 \oZ_{R \a}
 \oC_{\dot \a}
 Z_{R}^{\dot \a}
  C^{\a}
+\lt ( Z_R \xp \f_L 
+ Z_L \xp \f_R 
+ \U \xp H 
+ J \xp \h + *\rt ) 
\ee

\stepcounter{orange} \la{CHCMAction}
{\bf \theorange}.\;
{\bf \PB\ for \HCM}: 
We termporarily dropped the   indices $i$ in the above and now we will restore them to their proper places.
The Action above in \pgh\ \ref{CHCMAction} yields zero\footnote{ We have dropped the
Source $\S_{\a \dot\a}$ for the auxiliary Field here of course. This is an application of the theorem in \pgh\ \ref{theoremaboutaux}.}
for the following \PB\footnote{We have simply removed the terms $\fr{\d \cA}{ \d \ov\S_i^{\a\dot \a}}
\fr{\d \cA}{ \d W^i_{\a\dot \a}}
$ from \ref{phipb} in accord with the theorem in \pgh\ \ref{theoremaboutaux}.}:
\be
\cP_{\rm DHC}[\cA]=
\int d^4 x \lt \{
\fr{\d \cA}{ \d Z_L^{i\dot \a}}
\fr{\d \cA}{ \d \f_{Ri\dot \a}}
+
\fr{\d \cA}{ \d Z_{Ri}^{\dot \a}}
\fr{\d \cA}{ \d \f^i_{L\dot \a}}
+
\fr{\d \cA}{ \d \ov \U_i}
\fr{\d \cA}{ \d H^i}
+\fr{\d \cA}{ \d \ov J_i}
\fr{\d \cA}{ \d \eta^i }
+*
\rt \}
+\fr{\pa \cA}{ \pa h_{\a\dot \a} }
\fr{\pa \cA}{ \pa \x^{\a\dot \a} }
\la{phipb2}\ee

\stepcounter{orange} \la{majremark}
{\bf \theorange}. \; {\bf Remarkable Symmetries of  the Action  $\cA_{\rm DHC}:$}
As was remarked for the Majorana case in \pgh\ \ref{majremark} above, the above Dirac type Action $\cA_{\rm DHC}
$ has a remarkable symmetry which was not obvious before we integrated the auxiliary $W$. The Field $\h$ and the Source $\U$ appear in similar ways to each other.  The Field $H$ and the Source $J$ also appear in similar ways to each other. The term $\fr{1}{ 2 }
\pa_{\a  \dot \a} J
\pa^{\a  \dot \a}
\oJ$ looks like a kinetic term for $J$, except that $J$ is a Source.  The term 
$ \eta  \lt (
 \fr{1}{ \sqrt{2} } \ov\f_{L   \d}
C^{\d}
+\fr{1}{ \sqrt{2} }  
 \f_{R \dot\d}
\oC^{\dot \d}
\rt )
$ looks like a Zinn Source coupled to a variation, except that $\h$ is a Field.

\refstepcounter{orange}\label{diracgen1} 
{\bf \theorange}.\; This symmetry can be exploited with  the following Generator for an  \CTR\ of the Action which will leave the \PB\ invariant, but of a new form. Our new Action will have `new Fields' $(H_L, H_R)$ and `new Sources'
$(\G_L,\G_R)$.  These will replace  the `old Fields'  $(H ,\h)$
 and the `old Zinn Sources' $(J, \U)$.
We choose a generating function $\cG_{\rm DHC}$ of the new
Zinn Sources $(\G_L,\G_R)$ and the old Field $H$ and the old Zinn Source $J$:

\be
\cG_{\rm DHC} = 
\int d^4 x 
\lt \{
 \fr{1}{\sqrt{2}}\lt (\ov H_i + \oJ_{i}\rt) 
 \G_{L}^i
+
\fr{1}{\sqrt{2}}\lt (  J^i -H^i \rt)
\G_{Ri} 
 +
 \fr{1}{\sqrt{2}}\lt ( H^i+  J^i\rt)
\ov\G_{Li}
+
\fr{1}{\sqrt{2}}\lt (  \oJ_{i} -\ov H_i   \rt)
\ov \G_{R}^i 
 \rt \}
\ee

\refstepcounter{orange}
{\bf \theorange}.\;
{\bf Equations from the Generator and their inverses}
Now using the Generator defined in \pgh\ \ref{diracgen1} above, we get

\[
H_L^i=\fr{\d \cG}{\d \G_{Ri}}=
\fr{1}{\sqrt{2}}\lt ( J^i- H^i  \rt)
;\;H_{Ri}=\fr{\d \cG}{\d \G_{L}^i}=
 \fr{1}{\sqrt{2}}\lt (\ov H_i + \oJ_{i}\rt) 
\la{canR}
\]
\[
\ov \h_{i}=\fr{\d \cG}{\d J^i}=
\fr{1}{\sqrt{2}} 
\lt (
\G_{Ri} 
+ \ov \G_{Li} 
\rt )
;\; \ov \U_{i}=\fr{\d \cG}{\d G^i}=
\fr{1}{\sqrt{2}} 
\lt (-
\G_{Ri} 
+ \ov \G_{Li} 
\rt )
\]
\[
\G_{Ri}=
\fr{1}{ \sqrt{2} }\lt ( -\ov\U_{i}+   \ov\h_{i}\rt )
;\; \G_{L}^i=
\fr{1}{ \sqrt{2} }\lt (
 \U^i + \h^i \rt )
;\;
\]
\be
\ov J_{i}=
\fr{1}{ \sqrt{2} }
\lt (
H_{Ri} 
+ \ov H_{Li} 
\rt )
;\; \ov H_i=
\fr{1}{ \sqrt{2} }
\lt (
H_{Ri} 
- \ov H_{Li} 
\rt )
\ee

\refstepcounter{orange}\la{diracpairChiralmults}
{\bf \theorange}.\;{\bf Chiral Action for Dirac Multiplet}
Using the above definitions, the Action can be written in terms
of the new variables as follows:

\[
\cA_{\rm DCL}
= \int d^4 x \; \lt \{
-\f_{L}^{i\dot \a}
\pa_{ \a\dot\a}
\ov\f_{Li}^{ \a}
+\fr{1}{ 2 }
\pa^{ \a\dot\a} \ov H_{Li} 
\pa_{ \a\dot\a}
 H_L^i 
+
\G_{Ri} 
 \f^i_{L \dot \d}
\oC^{\dot \d}
+
 \ov\G_{R}^i  
 \ov\f_{L  i \d}
C^{ \d}
\rt.\]
\be
\lt.
+
 Z_{Ri}^{\dot \a}
  \pa_{ \a \dot \a} H_L^i
C^{\a} 
+ 
 \oZ_{R}^{ i\a}
  \pa_{ \a \dot \a} \ov H_{Li}
C^{\a} 
+
 Z_{R i\dot \a}
  C_{\a}
 \oZ_{R}^{ i \a}
  \oC^{\dot\a}
\;{\rm + \x\; terms}
\rt\}
\ee
plus a similar term
with $L \ra R$:
\[
\cA_{\rm DCR}
= \int d^4 x \; \lt \{
-\f_{Ri}^{\dot \a}
\pa_{ \a\dot\a}
\ov\f_{R}^{i \a}
+\fr{1}{ 2 }
\pa^{ \a\dot\a} \ov H^i_R 
\pa_{ \a\dot\a}
 H_{Ri} 
+
\G_L^i 
 \f_{Ri \dot \d}
\oC^{\dot \d}
+
 \ov\G_{Li}  
 \ov\f^i_{R  \d}
C^{ \d}
\rt.
\]
\be\lt.+
 Z_{L}^{i\dot \a}
  \pa_{ \a \dot \a} H_{Ri}
C^{\a} 
+
 \oZ_{Li}^{ \a}
  \pa_{ \a \dot \a} \ov H_R^i
C^{\a} 
+
 Z^i_{L \dot \a}
  C_{\a}
 \oZ_{Li}^{  \a}
  \oC^{\dot\a}
\;{\rm + \x\; terms}
\rt\} 
\ee
We must add $+ \cA_{\rm SUSY}$ as usual to the above.
This is a pair of \CM s that we can put together to make a Dirac \CM. 

\refstepcounter{orange} \la{MCPB}
{\bf \theorange}.\;
{\bf New Form of \PB} 
In terms of the new Fields, the 
\PB\ takes the following
`separated' form.
\be
\cP_{\rm   DCL}[\cA]
=
\int d^4 x \lt \{
\fr{\delta \cA}{ \delta Z_{Ri}^{\dot \a}}
\fr{\delta \cA}{ \delta \f^i_{L\dot \a}}
+
\fr{\delta \cA}{ \delta  \G_{Ri}}
\fr{\delta \cA}{ \delta H_L^i }
+
\fr{\delta \cA}{ \delta \oZ_{R}^{i  \a}}
\fr{\delta \cA}{ \delta \ov \f_{Li \a}}
+\fr{\delta \cA}{ \delta  \oG_R^i}
\fr{\delta \cA}{ \delta \ov H_{Li} }
\rt \} 
\la{Gsepcc}
\ee
plus
\be
\cP_{\rm   DCR}[\cA]
=
\int d^4 x \lt \{
\fr{\delta \cA}{ \delta Z_L^{i\dot \a}}
\fr{\delta \cA}{ \delta \f_{Ri\dot \a}}
+
\fr{\delta \cA}{ \delta  \G_L^i}
\fr{\delta \cA}{ \delta H_{Ri} }
+\fr{\delta \cA}{ \delta  \oG_{Li}}
\fr{\delta \cA}{ \delta \ov H_R^i }
+
\fr{\delta \cA}{ \delta \oZ_{Li}^{  \a}}
\fr{\delta \cA}{ \delta \ov \f^i_{R \a}}
\rt \}
\la{dsfsdgbgs}
\ee
We must add $+\fr{\pa \cA}{ \pa h_{\a\dot \a} }
\fr{\pa \cA}{ \pa \x^{\a\dot \a} }$ as usual to the above.
These add to make a  Dirac \PB\ for the \CM\ pair.

\tiny \articlenumber
\\
\today
\end{document}

%% file: Johnmacros21.tex
\newcommand{\pgh}{{Section}}
\newcommand{\Fpi}{Feynman path integral}

\newcommand{\UC}{Un-Chiral}
\newcommand{\UCM}{Un-Chiral Multiplet}
\newcommand{\ic}{\int d^4 x\; }

\newcommand{\etc}{{+\; {\rm etc.}}}

\newcommand{\xp}{{\x\cdot\pa}}

\newcommand{\SG}{{Supergravity}}

\newcommand{\CTR}{{Exchange\;Transformation}}

\newcommand{\HC}{{Half-Chiral}}

\newcommand{\CM}{{Chiral Multiplet}}

\newcommand{\HCM}{{Half-Chiral Multiplet}}

\newcommand{\MHCM}{{Majorana\;Half-Chiral Multiplet}}

\newcommand{\AF}{{Auxiliary Field}}

\newcommand{\ICDSS}{{Irreducible Chiral Dotted Spinor Superfield}}
\newcommand{\ICDSM}{{Irreducible Chiral Dotted Spinor Supermultiplet}}

\newcommand{\cG}{{\cal G}}

\newcommand{\cA}{{\cal A}}

\newcommand{\cdss}{Chiral Dotted Spinor Superfield}

\newcommand{\SYM}{SUSY Yang-Mills theory}
\newcommand{\CC}{Complex Conjugate}

\newcommand{\PB}{Master Equation}

\newcommand{\dB}{\d_{{}_{{}_{\rm BRST}}}}

\newcommand{\bt}{\begin{tabular}{c}}
\newcommand{\et}{\end{tabular}}

\newcommand{\eb}{\ee\be } 
\newcommand{\ebp}{\rt.\ee\be\lt.} 
\newcommand{\bmat}{\lt ( \begin{array} }
\newcommand{\emat}{  \end{array} \rt )}

\newcommand{\ovW}{{\ov W}}

\newcommand{\oH}{{\ov H}}

\newcommand{\oZ}{{\ov Z}}

\newcommand{\cP}{{\cal P}}

\newcommand{\oJ}{{\ov J}}

\newcommand{\oS}{{\ov S}}

\newcommand{\oG}{{\ov \G}}

\newcommand{\ul}{\underline}
\newcommand{\ED}{\end{document}}

\newcommand{\oY}{{\ov Y}}
\newcommand{\og}{{\ov g}}
\newcommand{\oy}{{\ov \y}}

\newcommand{\om}{{\ov m}}

\newcommand{\oC}{{\ov C}}

\newcommand{\oF}{{\ov F}}
\newcommand{\A}{{\ov A}}

\renewcommand{\a}{\alpha}	
\renewcommand{\b}{\beta}
\newcommand{\g}{\gamma}
\renewcommand{\d}{\delta}

\newcommand{\z}{\zeta}
\newcommand{\h}{\eta}

\newcommand{\lam}{\lambda}
\newcommand{\m}{\mu}
	
\newcommand{\x}{\xi}

\newcommand{\s}{\sigma}

\newcommand{\f}{\phi}
\renewcommand{\c}{\chi}
\newcommand{\y}{\psi}
\newcommand{\w}{\omega}
\newcommand{\G}{\Gamma}

\newcommand{\Lam}{\Lambda}

\renewcommand{\S}{\Sigma}
\newcommand{\U}{\Upsilon}

\newcommand{\la}{\label}
\newcommand{\ci}{\cite}

\newcommand{\ds}{\documentstyle}	
\newcommand{\fr}{\frac}

\newcommand{\pa}{\partial}
\newcommand{\ov}{\overline}
\newcommand{\be}{\begin{equation}}
\newcommand{\ee}{\end{equation}}
\newcommand{\ba}{\begin{array}} 
\newcommand{\ea}{\end{array}}
\newcommand{\bea}{\begin{eqnarray}}
\newcommand{\eea}{\end{eqnarray}}
\newcommand{\ra}{\rightarrow}

\newcommand{\lt}{\left}
\newcommand{\rt}{\right}

\newcommand{\ben}{\begin{enumerate}}
\newcommand{\een}{\end{enumerate}}
\newcommand{\bitem}{\begin{itemize}}
\newcommand{\eitem}{\end{itemize}}

%% file: 4933SuppessedSUSYFINALX.bbl
\begin{thebibliography}{99}

\bibitem{west}  Peter West, Introduction to Supersymmetry and Supergravity, World Scientific (1990).


\bibitem{superspace}  S. J. Gates, M. T. Grisaru, M. Rocek and W. Siegel, Superspace, Benjamin, 1983.

\bibitem{WB}J. Wess and J. Bagger,  Supersymmetry and Supergravity, Second Edition, Princeton University Press (1992).  
 
 


\bibitem{ferrarabook} Many of the original papers on  SUSY and Supergravity are collected in  Supersymmetry, Vols. 1 and 2, ed. Sergio Ferrara, North Holland, World Scientific, (1987).

\bibitem{KBbook}
Joseph Buchbinder, Sergio M. Kuzenko:
``Ideas and Methods of Supersymmetry and Supergravity"
Chapman \& Hall/CRC, Second Edition,
Series in High Energy Physics, Cosmology and Gravitation,Edition	2, 
ISBN	1584888644, 9781584888642

  \bibitem{Weinberg3} Steven Weinberg: ``The Quantum Theory of Fields" Volume 3, Cambridge University Press, ISBN 052155002. 


\bibitem{Gatesfundremains} 
  S.~J.~Gates, Jr., W.~D.~Linch, III, J.~Phillips and L.~Rana,
  ``The Fundamental supersymmetry challenge remains,''
  Grav.\ Cosmol.\  {\bf 8}, 96 (2002)
  [hep-th/0109109].

\bibitem{texts} A recent textbook introduction to AdS CFT   can be found in
\ci{Freedman:2012zz}, to Branes in \ci{newwest}, and to M theory in 
\ci{becker}.

\bibitem{Freedman:2012zz} 
  D.~Z.~Freedman and A.~Van Proeyen,
`Supergravity', Cambridge, UK: 
ISBN: 9781139368063 (eBook), 9780521194013 (Print)
  (Cambridge University Press)

 \bibitem{newwest}
A recent textbook is: Peter West,
Introduction to Strings and Branes,
Cambridge University Press, (2012)
        ISBN: 9780521817479

 \bibitem{becker}
 Katrin Becker, Melanie Becker, John H. Schwarz ``String Theory and M-Theory: A Modern Introduction'' Cambridge University Press, (2007)




\bibitem{xerxes} 
  H.~Baer and X.~Tata,
  ``Weak scale supersymmetry: From superfields to scattering events,''
  Cambridge, UK: Univ. Pr. (2006)  

\bibitem{haber}
  K.~A.~Olive {\it et al.}  [Particle Data Group Collaboration],
  ``Review of Particle Physics,''
  Chin.\ Phys.\ C {\bf 38}, 090001 (2014).
 The articles on page 1554 (Theory) and page 1573 (Experiment) give a nice summary of
models and experimental searches based on this working hypothesis. 





\bibitem{susymeets} There have been regular meetings every year for a long time with names like  ``SUSY 2016" which make contact between experiment and SUSY.   These are sometimes published and can be found on the Web.




\bibitem{one}
  J.~A.~Dixon,
  ``An Irreducible Massive Superspin One Half Action Built From the Chiral Dotted Spinor Superfield,''
  Phys.\ Lett.\ B {\bf 744}, 244 (2015)
  [arXiv:1502.07680 [hep-th]].
\bibitem{two}
  J.~A.~Dixon,
  ``An Extraordinary Mass Invariant and an Obstruction in a Massive Superspin One Half Model made with a Chiral Dotted Spinor Superfield,''
  Phys.\ Lett.\ B {\bf 749}, 153 (2015)
  [arXiv:1505.06838 [hep-th]].
 

\bibitem{three}  
  J.~A.~Dixon,
  ``SUSY Invariants from the BRST Cohomology of the SOSO Model,''
  arXiv:1506.02366 [hep-th].


\bibitem{johnsusy} These objects with uncontracted indices were discussed in  a long series of papers including 
\cite{Dixonholes}
\cite{DixonChiralcohom}
\ci{Dixon:1993jt}  \cite{DixonRahm} \ci{Dixonjumps} .  They were based on the paper 
\ci{Dixonspec}.  Spectral sequences are explained nicely in  \ci{Mcleary}.









\bibitem{Dixonholes} 
  J.~A.~Dixon,
  ``Supersymmetry is full of holes,''
  Class.\ Quant.\ Grav.\  {\bf 7}, 1511 (1990).

\bibitem{DixonChiralcohom} 
  J.~A.~Dixon,
  ``BRS cohomology of the Chiral Superfield,''
  Commun.\ Math.\ Phys.\  {\bf 140}, 169 (1991).

\bibitem{Dixon:1993jt} 
  J.~A.~Dixon and R.~Minasian,
  ``BRS cohomology of the supertranslations in D = 4",
  Commun.\ Math.\ Phys.\  {\bf 172}, 1 (1995)
  [hep-th/9304035].


\bibitem{DixonRahm} 
  J.~A.~Dixon, R.~Minasian and J.~Rahmfeld,
  ``Higher spin BRS cohomology of supersymmetric Chiral matter in D = 4,''
  Commun.\ Math.\ Phys.\  {\bf 171}, 459 (1995)
  [hep-th/9308013].

\bibitem{Dixonjumps} 
  J.~A.~Dixon,
  `SUSY Jumps Out of Superspace in the Supersymmetric Standard Model,'
  arXiv:1012.4773 [hep-th].








\bibitem{Dixonspec} 
  J.~A.~Dixon,
  ``Calculation of BRS cohomology with spectral sequences,''
  Commun.\ Math.\ Phys.\  {\bf 139}, 495 (1991).

\bibitem{Mcleary} Mcleary, John: ``A User's Guide to Spectral Sequences'',
2nd Edition (2001)
Cambridge University Press, 
ISBN 0-521-56141-8-ISBN 0-521-56759-9 (paperback)
 

\bibitem{poissonbrak} The \PB\ goes back to \ci{Becchi:1975nq} and Zinn-Justin's early contribution is set out in  a later textbook \ci{zinnbook}.  An early and pithy introduction was in \ci{taylor}. A bit of history and Zinn Justin's involvement is in 
\ci{Zinnarticle}.  A more recent treatment is in \ci{Weinberg2}.

\bibitem{Becchi:1975nq} 
  C.~Becchi, A.~Rouet and R.~Stora,
  ``Renormalization of Gauge Theories,''
  Annals Phys.\  {\bf 98}, 287 (1976).

\bibitem{zinnbook} J. Zinn-Justin, ``Quantum Field Theory and Critical Phenomena", Oxford Science Publications, Reprinted 1990. 


\bibitem{taylor}
  J.~C.~Taylor,
  ``Gauge Theories of Weak Interactions,''
  Cambridge 1976, 167p




\bibitem{Zinnarticle}
 A summary and some history can be found in  J.~Zinn-Justin,
  ``From Slavnov-Taylor identities to the ZJ equation,''
  Proc.\ Steklov Inst.\ Math.\  {\bf 272}, 288 (2011).



  \bibitem{Weinberg2} Steven Weinberg: ``The Quantum Theory of Fields" Volume 2, Cambridge University Press, ISBN 052155002. 


\bibitem{weinbergcosmo}
  S.~Weinberg,
  ``The Cosmological Constant Problem,''
  Rev.\ Mod.\ Phys.\  {\bf 61}, 1 (1989).



  \bibitem{wittenreally}
  E.~Witten,
  ``Is supersymmetry really broken?,''
  Int.\ J.\ Mod.\ Phys.\ A {\bf 10}, 1247 (1995)
  [hep-th/9409111].





\bibitem{five} John A. Dixon,
``The SSM with Suppressed SUSY Charge".  arXiv:1604.06396 [hep-th]


\bibitem{BV} I. A. Batalin and G.A. Vilkovisky, ``Gauge Algebra and Quantization", Physics Letters B 102.1 (1981): 27-31

\bibitem{algebrarefs} 
  R.~Haag, J.~T.~Lopuszanski and M.~Sohnius,
  ``All Possible Generators of Supersymmetries of the S Matrix,''
  Nucl.\ Phys.\ B {\bf 88}, 257 (1975).

\bibitem{goldstein} Herbert Goldstein, "Classical Mechanics" (Second Edition)
Addison Wesley ISBN 0-201-02918-9

\bibitem{LandauLifmechanics} L.D. Landau and E.M. Lifschitz, "Mechanics",  Pergamon Press (1960) 


\bibitem{Witten:1981nf} 
  E.~Witten,
  ``Dynamical Breaking of Supersymmetry,''
  Nucl.\ Phys.\ B {\bf 188}, 513 (1981).
  doi:10.1016/0550-3213(81)90006-7






\bibitem{dixonnucphys}
  J.~A.~Dixon,
  ``Field Redefinition and Renormalization in Gauge Theories,''
  Nucl.\ Phys.\ B {\bf 99}, 420 (1975). 




\end{thebibliography}
